\documentclass[a4,oneside,11pt]{article}
\usepackage[usenames, dvipsnames]{color}
\usepackage{natbib}
\usepackage{graphicx}
\usepackage[]{hyperref}

\usepackage{nameref}
\usepackage{mathptmx}

\usepackage{amsfonts, amssymb, amsthm}
\usepackage{eurosym,pifont}
\usepackage[svgnames]{xcolor}
\usepackage{pgfplots}
\pgfplotsset{width=2cm,compat=newest}
\pgfplotsset{small
}
\usepackage{adjustbox}
\usepackage{pdfpages}

\usepackage{threeparttable}
\usepgfplotslibrary{groupplots} 
\usetikzlibrary{pgfplots.groupplots} 
\usepackage{array}
\usepackage[margin=1.5in]{geometry}
\usepackage{array,makecell}
\usepackage{booktabs}

\usepackage{thmtools}
\usepackage{bbding}
\declaretheoremstyle[
  spaceabove=6pt, spacebelow=6pt,
  headfont=\normalfont\bfseries,
  notefont=\mdseries\bfseries,
  notebraces={(}{)},
  bodyfont=\itshape,
]{mystyle}
\declaretheorem[style=mystyle]{Hypothesis}

\newtheorem{res}{Result} 
\usepackage{ulem}
\usepackage[section]{placeins}

\usepackage{caption}
\usepackage{subcaption}

\usepackage{lscape}

\usepackage[super]{nth}

\usepackage{hyperref}

\makeatletter
\newenvironment{chapquote}[2][2em]
  {\setlength{\@tempdima}{#1}%
   \def\chapquote@author{#2}%
   \parshape 1 \@tempdima \dimexpr\textwidth-2\@tempdima\relax%
   \itshape}
  {\par\normalfont\hfill--\ \chapquote@author\hspace*{\@tempdima}\par\bigskip}
\makeatother

\usepackage[]{setspace}
\usepackage{comment}
\usepackage{soul}

\usepackage[hang]{footmisc}
\definecolor{darkred}{rgb}{0.55, 0.0, 0.0}
\usepackage{amsmath}
\DeclareMathOperator*{\argmax}{arg\,max}
\theoremstyle{plain}

\newtheorem{prop}{Proposition}

\begin{document}

\begin{titlepage}
\renewcommand*{\thefootnote}{\fnsymbol{footnote}}

 \begin{spacing}{1.5} 
 
\title{Decreasing Incomes Increase Selfishness}
 \end{spacing}
 
  \begin{spacing}{0.6} 
 \author{
 \Large{Nickolas Gagnon} \textcircled{r} \Large{Riccardo D. Saulle} \textcircled{r} \Large{Henrik W. Zaunbrecher} \thanks{
\footnotesize 
Acknowledgements: The order of authors is certified random (AEA Random Author Archive Code: mBjvN9oaNVE1); the proper short-form reference is therefore Gagnon \textcircled{r} al. Henrik W. Zaunbrecher is very grateful for financial support from the Gesellschaft f\"ur experimentelle Wirtschaftsforschung e.V. (GfeW) through the Heinz
Sauermann-F\"orderpreis zur experimentellen Wirtschaftsforschung. Nickolas Gagnon is also grateful for financial support from the Fonds de recherche du Qu\'ebec - Soci\'et\'e et culture. We thank Gergely Hajdu, Jona Linde, Ali Ozkes, Jan Potters, Arno Riedl, Pietro Salmaso, and participants at TIBER 2018 and in the MLSE seminar in Maastricht for helpful comments and suggestions. Our study was approved by the Maastricht University’s Behavioral and Experimental Economics Laboratory at a public ethics review and project proposal meeting.
Authors information:
Henrik W. Zaunbrecher (corresponding author): Maastricht University, Department of Economics, P.O. Box 616, 6200 MD Maastricht, The Netherlands, {\tt h.zaunbrecher@maastrichtuniversity.nl}; Nickolas Gagnon: Vienna University of Economics and Business, Department of Strategy and Innovation, Institute for Markets and Strategy,  Welthandelsplatz 1, 1020 Vienna, Austria, {\tt nickolas.gagnon@wu.ac.at}; Riccardo D. Saulle: University of Padova, DSEA, Via del Santo 33, Padua, Italy,  {\tt riccardo.saulle@unipd.it}.
}\vspace{-0.5cm}
}

 \end{spacing}
 
 \begin{spacing}{1} 
 
\vspace{-0.3cm}
\date{\today}

\maketitle

\end{spacing} 

 \begin{spacing}{1.5}

\vspace{-0.6cm}
\begin{abstract}
\vspace{-0.1cm}
{\small 
 \noindent We use a controlled laboratory experiment to study the causal impact of income decreases within a time period on redistribution decisions at the end of that period, in an environment where we keep fixed the sum of incomes over the period. First, we investigate the effect of a negative income trend (intra-personal decrease), which means a decreasing income compared to one's recent past. Second, we investigate the effect of a negative income trend relative to the income trend of another person (inter-personal decrease). If intra-personal or inter-personal decreases create dissatisfaction for an individual, that person may become more selfish to obtain compensation. We formalize both effects in a multi-period model augmenting a standard model of inequality aversion. Overall, conditional on exhibiting  sufficiently-strong social preferences, we find that individuals indeed behave more selfishly when they experience decreasing incomes. While many studies examine the effect of income inequality on redistribution decisions, we delve into the history behind one's income to isolate the effect of income \textit{changes}.
 
} 
\end{abstract}

{\small 
}
\vfill
\thispagestyle{empty}

\end{spacing} 
\end{titlepage}

\section{Introduction}
\label{sec:intro}

{\small
 \begin{spacing}{1.5}

\begin{chapquote}{Kahneman and Tversky (1979), Prospect Theory: An Analysis of Decision under Risk} \noindent  The carriers of value or utility are changes rather than final asset positions
\end{chapquote}

\end{spacing}
}

Heterogeneous income growth is a central issue of our times \citep{milanovic2016global,piketty2014capital}. In the United States between 1980 and 2004, pre-tax real incomes increased by 121\% for the richest 10\%, 42\% for the richest 10--50\%, 7\% for the richest 50--80\%, and decreased by 25\% for the poorest 20\% \citep{piketty2017distributional}.\footnote{Comparable pictures emerge in other countries, more moderate in Europe and more extreme in Asia \citep{alvaredo2017world}. Some of this heterogeneity is spatial: several major industrial hubs experienced striking declines in average household incomes since the 1970s, e.g., Buffalo ($-23\%$) and Detroit ($-35\%$) \citep{hartley2013urban}. Other differences in the evolution of incomes are based on education \citep{goldin2007long}, gender \citep{blau2017gender}, and ethnicity \citep{bayer2016divergent}.} Concurrently, although economists make understanding the determinants of redistribution a priority \citep{alesina2009preferences}, the possible consequences of income changes for redistributive decisions have not been isolated. Loss aversion to past income and aversion to unequal income trends could both make an individual more selfish. Whether this is indeed the case is difficult to establish because income changes are intertwined with potential confounds. For instance, a person whose income decreased from $70,000$ USD per year to $50,000$ USD while his or her peer group's income stays at $70,000$ USD faces: a negative income trend ($-20,000$ USD), a negative relative income trend ($-20,000$ USD), a new income level ($50,000$ USD) and a new inequality level relative to the peer group ($-20,000$ USD). Drawing on the tradition of controlled laboratory experiments isolating the influence of income inequality on redistribution (e.g., \citealp{fehr1999theory,bolton2000erc}), our study offers the first evidence that income decreases indeed affect redistribution decisions, even after controlling for one's absolute income and for income inequality.

We designed a laboratory experiment, detailed in Section~\ref{sec:design}, in which individuals complete real-effort tasks in several distinct periods corresponding to our treatments. Every period contains two sub-periods, in each of which participants receive an exogenously-assigned income that we call ``wage" for completing an individual task (``wage" denotes a participant's income in a sub-period). The treatments that we implement vary the intra-personal and inter-personal income changes faced by two matched participants in a period by changing the wages between the first and second sub-periods. Furthermore, participants' income is taxed over the period and, at the end of the period, each of the two participants individually decides how the money deducted from both participants should be redistributed among them. We then implement one redistribution decisions per matched pair of participants. Crucially, each treatment manipulates income trends within a period while keeping constant the sum of incomes over that period before redistribution takes place at the end of the period.

Our contention that income decreases can influence redistribution decisions is rooted in the large number of studies documenting reference dependence, loss aversion, and inequality aversion. First, the literature on inequality aversion \citep{fehr1999theory,bolton2000erc} posits theoretically and shows empirically that individuals dislike that their income differs from the inter-personal reference point formed by the income of other individuals, especially falling behind the income of others.\footnote{In this paper, we focus on disadvantageous inequality aversion for two reasons. First, disadvantageous inequality aversion is assumed to be stronger in these models and a large number of empirical studies have gathered evidence supporting its existence. Second, our research is closely linked to research on loss aversion, which considers that individuals are especially sensitive to losses. The literature on inequality aversion is quite rich, and includes empirical studies using observational methods \citep{CLARK1996359, solnick1998more, luttmer2005neighbors}, natural experiments \citep{kuhn2011effects,card2012inequality}, field experiments \citep{cohn2014social,breza2017morale,dube2018fairness} as well as laboratory experiments \citep{fehr1999theory,bolton2000erc} and experiments with the general population \citep{bellemare2008measuring}.} They redistribute from themselves to others when their income is larger than others', and, especially, from others to themselves when their income is lower. It is possible that other inter-personal reference points exist, such as the change in income experienced by other individuals or, in other words, other individuals' income trend. Individuals would then take more from others to avoiding experiencing a disadvantageous change in their income relative to the change in others' income, i.e., to avoid facing a negative relative trend. The literature on inequality aversion often studies consequences of this aversion in other domains where income changes and income inequality are weaved together. For instance, \cite{card2012inequality}, \cite{cohn2014social}, \cite{breza2017morale}, and \cite{dube2018fairness} all study the effect of learning about wage inequality on labor decisions, where a worker faces simultaneously a change in believed or actual income inequality and a new level of inequality.\footnote{Another example is \cite{kuhn2011effects}, which analyzes changes in consumption after one's neighbor wins at a lottery. This situation entails a change as well as a new level of local income inequality.} Nevertheless, inequality aversion studies are not crafted to isolate consequences of individuals comparing themselves to others in terms of income changes, notably for redistribution decisions.

Second, research on reference-dependent preferences has long modeled that individuals experience a disutility when falling behind an intra-personal reference point \citep{kahneman1979prospect,tversky1991loss,kHoszegi2006model}. Most relevant to our study is the empirical research suggesting that the behavior of workers is consistent with a dislike for falling behind their own past income \citep{dellavigna2017reference,cohn2015}.\footnote{There is also evidence for other forms of reference dependency, such as workers disliking to fall behind their earnings goals \citep{camerer1997labor,crawford2011new} and behind their earnings expectations \citep{mas2006pay,abeler2011reference}.} \citet{loewenstein1991workers} also reports that individuals prefer increasing wage profiles to equivalent decreasing ones.\footnote{The same effect has been shown for other contexts such as experiences \citep{ross1991evaluations}, environmental outcomes \citep{guyse2002valuing}, and health \citep{chapman1996expectations}.} In a similar vein, self-reported well-being is lower when one's living standard decreases over time \citep{clark2008relative,senik2009direct}. In terms of redistribution decisions, if individuals take their own past income as their intra-personal reference point, it is conceivable that they redistribute more from others to themselves to avoid as much as possible to fall behind their past income or, in other words, to avoid or decrease the extent of a negative income trend. Nevertheless, none of the current studies analyzes the impact of income decreases on redistribution decisions.

Post experiment, we formalize in Section~\ref{sec:model} our intuitions regarding the effect of changing income trends on redistribution. We combine two prominent models in a multi-period setting: (i) the inequity aversion of \citet{bolton2000erc}, which is a one-period model, and (ii) the reference dependence of \citet{kHoszegi2006model}, in the form employed by \citet{dellavigna2017reference}, among others, which is loss aversion relative to previous income. Our model considers two types of inequality aversions: inequality aversion in income and inequality aversion in income trends. That is, we assume that individuals not only care about relative incomes, but also about relative income trends. Furthermore, we assume that they are averse to a negative income trend, meaning that they suffer a specific utility loss when their incomes are lower than their past incomes. In addition to providing a more solid underpinning for our pre-experiment predictions (Section~\ref{sec:hypo}), the model offers an additional testable insight: decreasing incomes can only negatively affect the generosity of individuals who are at least minimally generous in the absence of income trends. This is important because most participants in our experiment never share anything. The reason is straightforward: fully selfish individuals have no scope for becoming more selfish (and almost fully selfish ones have little scope for doing so).

Key to our investigation is the causal relationship that we obtain between wage decreases and redistribution decisions, which is provided by the laboratory experiment. This methodology offers two main advantages. First, we randomly assign income trends, which precludes that individuals experiencing different trends do so because of different underling characteristics. Second, crucially, we can study income changes within a time period without changing the sum of incomes and income inequality over that period. That is, we clearly separate the role of  income decreases from the role of absolute income and the role of disadvantageous income inequality.

Our results, provided in Section~\ref{sec:results}, are as follows. For the sample as a whole (including selfish individuals), we report that, at a given absolute income and inequality level, there is a negative effect of decreasing incomes on generosity, but the effect is significant only at marginal levels (10$\%$). In line with our model, we find strong evidence of this negative effect for individuals with social preferences (i.e., those who exhibit sufficiently-strong disadvantageous inequality aversion). The overall effect is significant for all socially-inclined participants considered together. When we consider participants in the two roles of the experiment separately, the effect is significant for those who earn relatively more (High Earners), although, in our preferred robust specification, it is not for those who earn relatively less (Low Earners). Importantly, this is not dependent on the specific cutoff point that we use for defining participants with social preferences. When it comes to the types of decreases, inter-personal decreases have a significant effect on the combined participants as well as High and Low Earners considered separately, and the effect of strictly intra-personal decreases is generally not significant. There is evidence that inter-personal decreases have a stronger effect than strictly intra-personal decreases, but the difference is only significant at marginal levels. 

Even though the time span of a period in our experiment is quite limited (8 minutes), the estimated average effect of income decreases is economically substantial for socially-inclined participants: they share approximately $12\%$ (1/4 of a standard deviation) less with their matched participant. Given that previous studies show that elicited social preferences map to political support for redistributive policies (e.g., \citealp{fisman2017distributional,kerschbamer2020social,epper2020other,almaas2020cutthroat}), our results suggest that, at least for inequality-averse individuals, declining absolute or relative incomes could contribute to the support for or opposition to redistributive policies.\footnote{While the degree of inequality aversion is moderate in our experiment, student samples provide a lower bound for the extent of inequality aversion in the population \citep{snowberg2021testing}. Social preferences elicited in experiments have also been shown to correlate with behavior outside of those experiments in other domains, such as loan repayments \citep{karlan2005using}, donations and other pro-social behaviors \citep{benz2008people,baran2010can,franzen2013external}, work productivity \citep{cohn2015}, and socially-responsible investments \citep{riedl2017investors}.} We conclude the paper with a brief discussion in Section~\ref{sec:conclusion}.

\section{Changing Income Trends and Measuring Redistribution}
\label{sec:design}

We designed an experiment consisting of five periods. In a period, two participants are anonymously matched together. A period consists of two sub-periods in each of which participants perform a real-effort task for a wage and it ends with a redistribution decision. Note that we employ the term ``wage" to refer to a participant's income from a sub-period.

\begin{figure}[h!]
\centering 
\includegraphics[width=\textwidth]{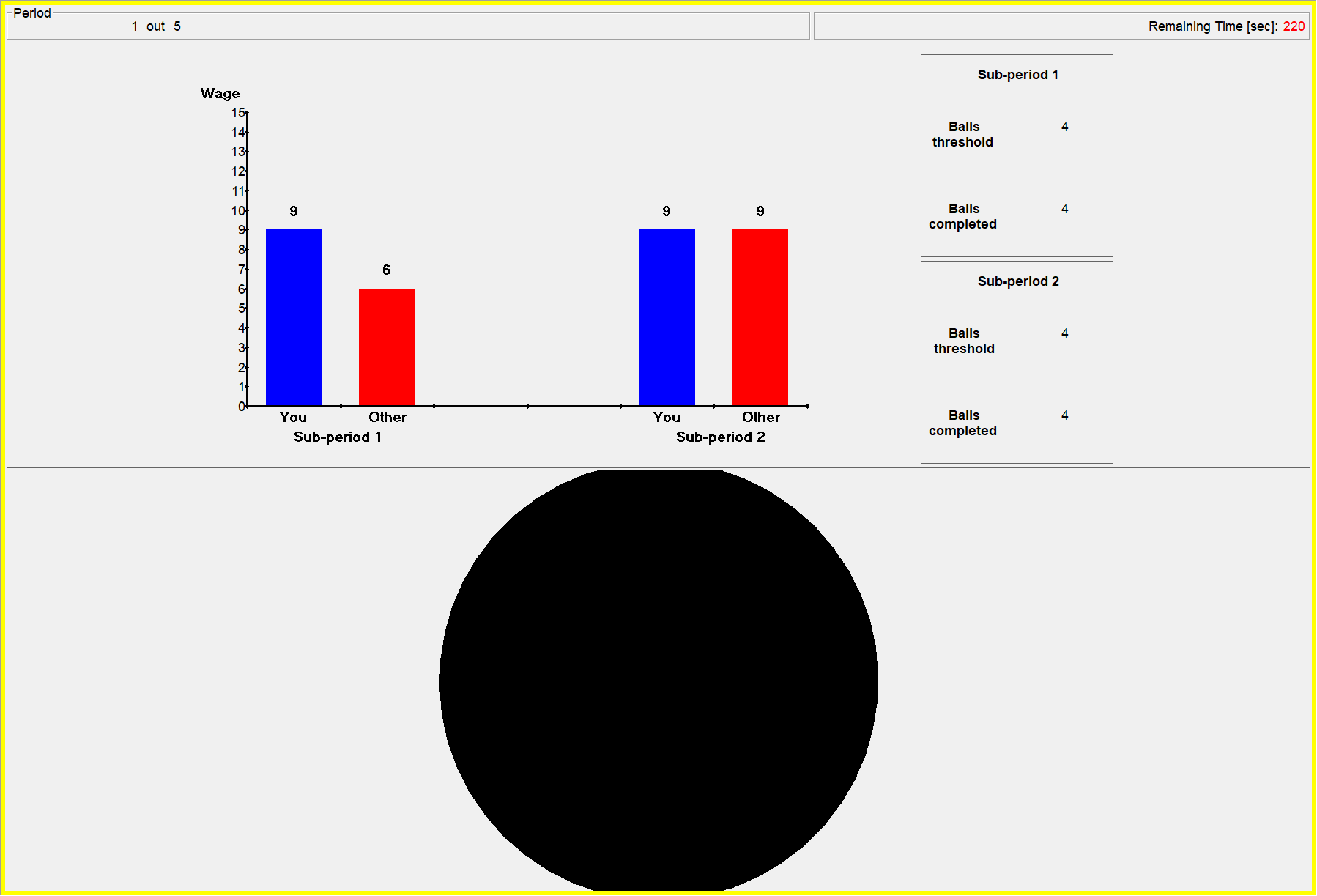}
 \caption{Screenshot of
 task in sub-period 2 (\textsc{CatchingUp} treatment)}\label{fig:example_task}
\end{figure}

The effort task in every sub-period is to reduce the size of four circles on the computer screen until they disappear. This is done by repeatedly clicking on a circle with the mouse while it moves across the screen. Only one circle appears at the time, and each click on it slightly decreases its size. A new circle appears once a circle completely disappears.\footnote{Participants are not paid more if they complete more than four circles. The task is a modified version of the one in \citet{cacault2016group}.} Participants have four minutes to complete the task, which can be completed easily by exerting a reasonable effort (most participants take approximately two minutes to finish the task). They are provided with a countdown and with a record of how many circles they have completed so far. Figure~\ref{fig:example_task} provides a screenshot of the task as experienced by participants in sub-period 2. Note that, in the experiment, we call circles ``balls" and that we provide participants with the reminder that they need to reach a ``Ball Threshold" of four to indicate that they need to make four circles disappear in order to earn the wage.

At the start of a period, participants are informed of their wage for the first sub-period a few seconds before the first sub-period starts. During a period, they monitor how their own wage and the wage of their matched partner evolve over the two sub-periods. This information is visualized through one graph exhibiting one's own wages and the wages of the other participant over the period up to the current sub-period. That is, participants see the wages in the current sub-period, and if they are in the second sub-period, they also see the wages from the previous sub-period. In addition to providing a screenshot of the task, Figure~\ref{fig:example_task} also shows how participants observe the evolution of wages in sub-period 2. The screen presents the wage information from the first sub-period on the left part of the graph, and from the second sub-period on the right part of the graph. Participants have an additional minute to rest in between the two sub-periods. A few seconds before the second sub-period starts, they are informed about any wage changes that occur between the sub-periods.

The participants are paid the two wages of a period---one for each of the two sub-periods---if and only if they complete the task in both sub-periods.\footnote{A participant is paid nothing for a period if the task in one of the two sub-periods is not completed. However, we set the wage high enough relative to the effort required for the task so that this only affected 2 out of 298 participants. We excluded these participants and their matched participants because the matched participants could see that those did not complete the task. Therefore, a total of four participants were dropped for the data analysis.} A third of each wage that they earn during the two sub-periods is taken from them and placed in a joint account. That is, the joint account contains a third of the two wages of one participant, and a third of the two wages of the other participant.\footnote{To ease the explanation for participants, we phrase the parts of the income taken as taxes and the distribution decisions as a redistribution of taxes collected. We collect a fix percentage of income to make it easier for participants to understand and calculate how much was taken from them.}

At the end of each period, the two participants individually propose how to distribute the money contained in the joint account, which has been collected from their wages for the two sub-periods. This is implemented through a dictator game with role uncertainty. That is, one of the two choices is randomly chosen to count. The money in the joint account always amounts to 11 Euro. Participants can keep the entire joint account for themselves, transfer its content to the other participant, or chose any in-between allocation in increments of 10 cents.\footnote{Figure~\ref{fig:example_redistribution} in Appendix~\ref{APP:B} provides a screenshot showing how the redistribution decision is presented to participants.} To further distinguish the different periods, participants then take a two-minute break before the next period starts. At the end of the experiment, one of the two participants' choices from one period is randomly chosen to count for payment.

We chose a within-subject design in order to increase statistical power \citep{bellemare2014statistical}. We employ five treatments that we implement for each participant over the five periods. The treatments vary the wage changes faced by participants between the first and second sub-period of a period. A participant is always in the role of either the High Earner or the Low Earner. Participants are not informed that they stay in their role. In the first role (High Earner), a participant always experiences advantageous income inequality over the period. That is, the sum of the two wages in a period is always 18 Euro. Similarly, the Low Earner always faces disadvantageous income inequality over the period---the sum of the two wages is always 15 Euro. This allows us to maintain the same income inequality over the period in all treatments, such that income inequality cannot explain any treatment differences.\footnote{We found it natural to create income inequality over each period because income inequality is common outside of the laboratory. A possible alternative design is to impose income equality over the period in all treatments. The hypotheses presented in Section~\ref{sec:hypo} would be the same under that alternative.
}

\begin{table}\caption{Overview of Income Trends in Treatments} 
\centering
\bigskip
\adjustbox{max width=\textwidth}{
\begin{threeparttable}

\begin{tabular}{ >{\centering\arraybackslash}m{7cm}
>{\centering\arraybackslash}m{4.5cm}}
\hline \hline 
  Treatment &\textcolor{Blue}{High Earner} / \textcolor{OrangeRed}{Low Earner} \\
  \hline   
\textsc{Stable}   &
\begin{tikzpicture}[baseline]
\begin{axis}[domain=0:2,ymin=0,ymax=15,width=4.5cm,  xtick={1,5},
    xticklabels={1,2},enlarge x limits=0.9 ]
    \addplot[ybar,fill=Blue] coordinates {
        (0.5,9)
        (4.5,9)
        
    };
      \addplot[ybar,fill=OrangeRed] coordinates {
        (1.5,7.5)
        (5.5,7.5)
    };  
\end{axis}
\end{tikzpicture}
\\ \hline
\textsc{IntraDecrease}

\footnotesize{(Intra-personal decrease for Low and High Earners)}

&
\begin{tikzpicture}[baseline]
\begin{axis}[domain=0:2,ymin=0,ymax=15,width=4.5cm, xtick={1,5},
    xticklabels={1,2},enlarge x limits=0.9 ]
    \addplot[ybar,fill=Blue] coordinates {
        (0.5,13.5)
        (4.5,4.5)
    }; \addplot[ybar,fill=OrangeRed] coordinates {
        (1.5,12)
        (5.5,3)
    };
\end{axis}
\end{tikzpicture}

\\ \hline
\textsc{IntraIncrease}

\footnotesize{(Intra-personal increase for Low and High Earners)} &
\begin{tikzpicture}[baseline]
\begin{axis}[domain=0:2,ymin=0,ymax=15,width=4.5cm, xtick={1,5},
    xticklabels={1,2},enlarge x limits=0.9 ]
    \addplot[ybar,fill=Blue] coordinates {
        (0.5,4.5)
        (4.5,13.5)
    };
        \addplot[ybar,fill=OrangeRed] coordinates {
        (1.5,3)
        (5.5,12)
    };
\end{axis}
\end{tikzpicture}

\\ \hline
\textsc{IntraInterChange}

\footnotesize{(Intra- and inter-personal decrease for Low Earners)}

&
\begin{tikzpicture}[baseline]
\begin{axis}[domain=0:2,ymin=0,ymax=15,width=4.5cm, xtick={1,5},
    xticklabels={1,2},enlarge x limits=0.9 ]
    \addplot[ybar,fill=Blue] coordinates {
        (0.5,4.5)
        (4.5,13.5)
    };    \addplot[ybar,fill=OrangeRed] coordinates {
        (1.5,12)
        (5.5,3)
    };
\end{axis}
\end{tikzpicture}

\\ \hline
\textsc{CatchingUp}

\footnotesize{(Inter-personal decrease for High Earners)}

&
\begin{tikzpicture}[baseline]
\begin{axis}[domain=0:2,ymin=0,ymax=15,width=4.5cm, xtick={1,5},
    xticklabels={1,2},enlarge x limits=0.9 ]
    \addplot[ybar,fill=Blue] coordinates {
        (0.5,9)
        (4.5,9)
    };    \addplot[ybar,fill=OrangeRed] coordinates {
        (1.5,6)
        (5.5,9)
    };
\end{axis}
\end{tikzpicture}
\\ \hline

\end{tabular}

\begin{tablenotes}
\item \footnotesize Wages are indicated on the vertical axis in Euro, and the two sub-periods of a period are indicated on the horizontal axis. Income inequality is constant over the period: income is 15 Euro for the Low Earner and 18 Euro for the High Earner. 
\end{tablenotes}
\end{threeparttable}
}

\end{table}

Table 1 details the wages of Low Earners and High Earners over the two sub-periods in the five treatments. The order of treatments is randomized.\footnote{There are 120 possible orders ($5!$). We overly sample from a random subset of the orders due to a software problem. That is, approximately 70\% of the orders are randomly drawn from a random subset of 30 orders---the subset itself is a random selection from the 120 orders---and the remaining 30\% is randomly drawn from the 90 other orders. Conducting the data analysis separately for each of those two sub-samples qualitatively provides the same results.} As we are interested in studying individual responses to absolute and relative wage decreases, we designed the following treatments varying wages changes. While those treatments do not cover all possible wage changes, they do provide several types of declining wages, as well as one set of stable wage and one increasing wage profile.

In \textsc{Stable}, the wage of each participant remains constant in the two sub-periods. In \textsc{IntraDecrease} and \textsc{IntraIncrease}, both participants face either an absolute wage increase or decrease. This allows us to study the effects of intra-personal wage changes. In \textsc{IntraInterChange}, the Low Earner experiences a wage decrease while the High-Earner experiences a wage increase. The wage changes are therefore both absolute and relative for the two participants. In \textsc{CatchingUp}, the wage of the Low Earner increases, while the wage of the High Earner is constant. That is, the High Earner encounters a relative wage decrease---the Low Earner is ``catching up" with the High Earner.

Before starting the experiment, the experimenter reads the instructions aloud and participants are provided with a written copy detailing all steps of the experiment.\footnote{We provide the original instructions in Appendix~\hyperref[APP:A]{A}.} After reading the instructions, participants complete comprehension questions, and help is provided if needed. They also go through a practice period, which is a shorter version of a real period, so that they become familiar with the proceedings of a period. This practice period includes the task and the redistribution decision, but does not count for payment. In it, each participant has the same wage, which stays constant of the two sub-periods. Participants are informed that they are paired with a participant in the same laboratory session in each period.
 
We designed the experiment using the software z-Tree \citep{fischbacher2007z}. It was conducted at the BEElab (Behavioural and Experimental Economics Laboratory) of Maastricht University in May-October 2019. Our sample consists of 294 participants recruited over 16 sessions via the online recruitment software ORSEE \citep{greiner2004}.\footnote{See footnote 6 explaining that four additional participants are not counted in our sample.} The experiment lasted for about 90 minutes and participants earned 16.50 Euro on average. 

\section{The Model}
\label{sec:model}

We build our model by combining (i) the inequity aversion model of \citet{bolton2000erc} (henceforth BO) and (ii) the reference dependence model of \citet{kHoszegi2006model} (henceforth KR), in the form employed by \citet{dellavigna2017reference}, which is loss aversion to previous income. Before we describe the model, transparency requires us to mention that, although we sketched its foundations for advancing pre-experiment predictions in Section~\ref{sec:hypo}, we only formalized it after conducting the experiment.\footnote{See \citet{bolton2005fair} for an example of post-hoc formalization.} Therefore, we do not test this model, but we rather employ it to better organize and enrich our ideas regarding the effect of income trends on redistribution decisions. Importantly, the model predicts that decreasing income trends should only affect the subset of individuals who are sufficiency altruistic (Proposition 2), a proposition that we subsequently use in Section~\ref{sec:results}.

We consider the following multi-period model with two individuals. In each period, each individual receives an income. We denote by $y_i\in \mathbb{R}_+$ the final income gained by individual $i$ over all periods, and by $t_i\in \mathbb{R}_+$ the individual's final income trend generated by the streams of wages over the periods. In particular, $t_i$ takes negative values when  the trend is decreasing and positive values otherwise. We assume that the individual has personal concerns, specifically (i) final income over the periods and (ii) final income trend, and social concerns, specifically (iii) relative final income over the periods and (iv) relative final income trend. Formally, individual $i$'s utility consists of four components: (i) {\it material payoff motivation function} $u_i(y_i)$, (ii) {\it trend gain-loss function} $\mu_i(t_i)$, (iii) {\it relative income motivation function} $\sigma_i(y_i,y_j)$, and (iv) {\it relative trend motivation function} $\tau_i(t_i, t_j)$. 
We describe the individual's utility as
\begin{align}
    U_{i}(u_i,\mu_i,\sigma_i, \tau_i).
\end{align}

We impose the restriction that (1) is additive and separable and that it satisfies the following assumptions:
\bigskip

\begin{description}

\item{\bf A0:} {\it $u_i(x)$, $\sigma_i(x)$ and $\tau_i(x)$,  are  continuous and twice differentiable for all $x$}.

\item{\bf  A1:} {\it  $\mu_i(t_i)$ is continuous, differentiable for  $t_i\neq 0$ with $\mu_i(0)=0$, and strictly increasing}.

\item{\bf A2:}
{\it If $z>x>0$, then $ \mu(z)+\mu(-z)<\mu(x)+\mu(-x)$}.

\item{\bf  A3:} {\it $\mu_i(t_i)$ is  concave for  $t_i>0$ and  convex for $t_i<0$}.

\item{\bf  A4:}  $\lim_{t\to 0}\frac{\mu'(-|t_i|)}{\mu'_i(|t_i|)}\equiv \lambda>1$.

\item{\bf  A5:} {\it $u_i(y_i)$ is increasing and concave  in $y_i$}.

\item{\bf A6:} 
{\it $\sigma_i$ is increasing and concave in $y_i$ with a maximum  at $y_j$}.

\item{\bf A7:} 
{\it $\tau_i$ is increasing and concave in $t_i$ with a  maximum  at $t_j$}.

\end{description}

We make assumption {\bf A0} for mathematical convenience. Assumptions {\bf A1--4} recast {\it reference dependence} and {\it  loss aversion} as in KR. Finally, assumption {\bf A5--7} are in the spirit of {\it narrow self interests} and {\it comparative effect} in BO.

In our application, the material payoff motivation function is represented by
\begin{align*}
    u_i(y_i)=y_i
\end{align*}
which  satisfies {\bf A0} and {\bf A5}.
  In line with KR and thus with assumption {\bf A1--4}, we specify  the trend gain-loss function as
\begin{align*}
    \mu_i(t_i)=\begin{cases}
    \eta t_i & \text{if}\ \ t_i<0\\
     0 & \text{if}\ \ t_i\geq 0
    \end{cases}
\end{align*}
with $\eta\in [0,1)$.
  Next, we write the relative income and trends  motivation functions as follows
\begin{align*}
    \sigma_i(y_i,y_j)=\left(y_i-\frac{1}{2}(y_i+y_j) \right)^2
\end{align*}
\begin{align*}
    \tau_i(t_i,t_j)=\left(t_i-\frac{1}{2}(t_i+t_j) \right)^2
\end{align*}
where we take $\frac{1}{2}(y_i+y_j)$ and $\frac{1}{2}(t_i+t_j)$ to be the reference points, which is similar in form to BO and satisfies assumptions {\bf A0} and {\bf A5--7}.
Finally, we consider the following utility specification
\begin{align}
    U_i(y_i,y_j,t_i,t_j)=\begin{cases}
a_i(y_i+\eta t_i)-b_i\left[\left(\frac{1}{2}(y_i-y_j)\right)^2 + \left(\frac{1}{2}(t_i-t_j)\right)^2
    \right] & \text{if} \ \ t_i<0\\
    a_iy_i-b_i\left[\left(\frac{1}{2}(y_i-y_j)\right)^2 + \left(\frac{1}{2}(t_i-t_j)\right)^2
    \right] & \text{if}\ \ t_i\geq 0
        \end{cases}
\end{align}

where $a_i,b_i>0$ are the agent's sensitivity to personal concerns and social concerns, respectively. We interpret $a_i/b_i$ as the agent's type, i.e., the ratio
of weights attributed to the personal
and social components of the motivation function.\footnote{BO's $a_i/b_i$ is the ratio of weight placed on pecuniary and relative components, but the motivation function excludes our trend gain-loss function and our relative trend motivation function.} Note that our model reduces to a BO-type motivation function when $t_i=t_j=0$. Furthermore, the first two components of $U_i(u_i,\mu_i,\sigma_i, \tau_i)$ replicate the reference-dependence framework of KR. Specifically, the reference point in one's own trend is imposed to be zero in $\mu_i$, i.e., the individual experiences an extra disutility from his/her own trend being negative, while there is no extra utility from one's trend being positive. This is similar to \citet{dellavigna2017reference}, which assumes loss aversion to previous income. Our utility function therefore reduces to a KR-type function when a participant does not care about social comparisons ($b_i=0$)

\subsection{The Dictator Game}
\label{subsec:dictator}

Our experimental design boils down to a Dictator Game at the end of an economy with 2 sub-periods\footnote{For this specific application to the situation that we study with our experiment, we call each time period a sub-period instead of a period to be consistent with the wording in the rest of the paper. A period in the experiment consists of two sub-periods over which a treatment is administered and at the end of which a redistribution decision is made.}. This is therefore the game that we study here. At end of each sub-periods, each individual receives a wage. Let $W_i>0$ be individual $i$'s sum of wages over the two sub-periods. We denote by $\Delta_i$ the difference between the wage in the second sub-period and the wage in the first sub-period. In the Dictator Game, individual $i$, who plays the role of the dictator, chooses the share $1-s$ ($0 \leq s \leq 1$) of a tax account $T > 0$ to give to another individual $j$ who is in the role of the recipient and does not make any decisions ($T$ is independent of everything else). This additional amount modifies the wage in the second sub-period, which affects income levels as well as income trends. In light of our model, the final income of the dictator over the two periods is therefore given by $y_i=W_i+s$ and the final income trend by $t_i=\Delta_i+s$. Note that we use the term ``final" because we are concerned about the income and the income trend at the very end of the timeframe, i.e., after the redistribution decision. For the recipient, the final income and final income trend are given by $y_j=W_j+1-s$ and $t_j=\Delta_i+1-s$. Therefore, the utility function $U_i^D(y_i,y_j,t_i,t_j)$ of the dictator takes the form

\begin{align*}
    U^D_i(.)=\begin{cases}
         a_i(W_i+sT+\eta( \Delta_i+sT))\\
        \qquad \quad -b_i\left[\left(\frac{1}{2}(W_i-W_j+2sT-T)\right)^2 + \left(\frac{1}{2}(\Delta_i-\Delta_j+2sT-T)\right)^2
    \right]
   & \text{if} \ \ \Delta_i+sT<0\\
    a_i(W_i+sT)\\
    \qquad  \quad -b_i\left[\left(\frac{1}{2}(W_i-W_j+2sT-T)\right)^2 + \left(\frac{1}{2}(\Delta_i-\Delta_j+2sT-T)\right)^2
    \right] & \text{if}\ \ \Delta_i+sT\geq 0
        \end{cases}
\end{align*}

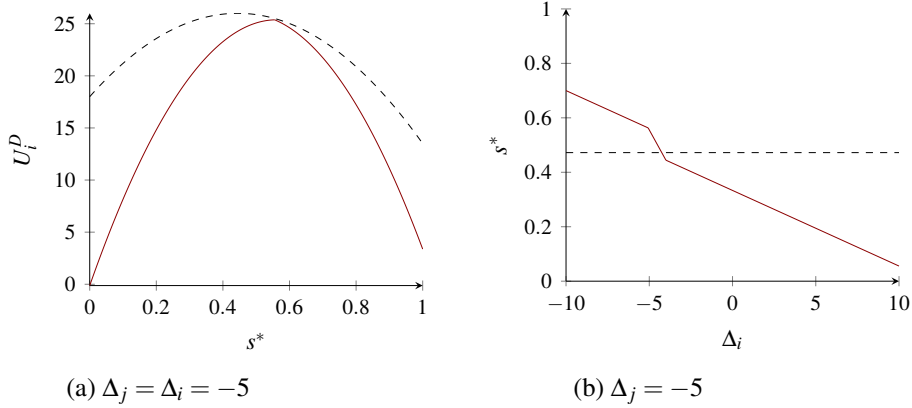
\begin{figure}[h]
\hspace{-2cm}
\centering
\begin{minipage}[t]{.3\textwidth}
\begin{tikzpicture}[thick, scale=0.9]
\begin{axis}[
    axis lines = left,
    xlabel = $s^*$,
    ylabel = {$U^D_i$},
]
\addplot [
    domain=0.555:1, 
    samples=100, 
    color=darkred,
]
 {2*(10+9*x)-(0.5)*(((0.5)*(18*x-4))^2+((0.5)*(18*x-9))^2)};

\addplot [
    domain=0:0.555, 
    samples=100, 
    color=darkred,
    ]
    {2*(10+9*x+(0.8)*(-5+9*x))-(0.5)*(((0.5)*(18*x-4))^2+((0.5)*(18*x-9))^2)};

\addplot [dashed,
    domain=0:1, 
    samples=100, 
    color=black,
    ]
    {2*(10+9*x)-(0.5)*(((0.5)*(18*x-4))^2} ;

\end{axis}
\end{tikzpicture}
\subcaption{$\Delta_j=\Delta_i=-5$}
\end{minipage}
\hspace{2cm}
\begin{minipage}[t]{.3\textwidth}
\begin{tikzpicture}[thick, scale=0.9]
\begin{axis}[
    axis lines = left,
    xlabel = $\Delta_i$,
    ylabel = {$s^*$},
       ymin=0,
    ymax=1
]

\addplot [dashed,
    domain=-10:10, 
    samples=100, 
    color=black,
]
{4/36+0.5-(5)/36};

\addplot [
    domain=-4:10, 
    samples=100, 
    color=darkred,
]
{4/36+0.5-(5)/36+(-5-x)/36};

\addplot [
    domain=-5.066:-4, 
    samples=100, 
    color=darkred,
]
{-x/9};

\addplot [
    domain=-10:-5.066, 
    samples=100, 
    color=darkred,
]
{4*(1.8)/36+0.5-(5)/36+(-5-x)/36};

\end{axis}
\end{tikzpicture}
\subcaption{$\Delta_j=-5$}    
\end{minipage}
    \caption{Examples of $U^D_i$ and $s^*$ for $a_i/b_i=4$, $W_i=10$, $W_j=5$, $T=9$, and $\eta=0.8$. The thick line indicates our model predictions and the dashed line indicates BO predictions. }
    \label{fig:example_dg_utility_sharing}
\end{figure}

The thick red curve in Figure~\ref{fig:example_dg_utility_sharing} panel (a) presents an example of the dictator's utility function, $U^D_i$, for our model. In contrast, the dashed black curve in the same panel illustrates the BO-style model, where we remove our trend gain-loss function and our relative trend motivation function. Figure~\hyperref[fig:example.add.utility]{C.1} in Appendix~\ref{APP:C} provides additional examples of $U^D_i$ for different agent types.

The  following proposition provides  optimality conditions for $U_i^D(y_i,y_j,t_i,t_j)$.

\begin{prop}

Define $\mathbb{H}\equiv W_i-W_j-2T-3\Delta_i-\Delta_j$. Then the function $U_i^D(y_i,y_j,t_i,t_j)$  is maximized at $s^*$ such that: 
\begin{align*}
      \argmax U_i^D(y_i,y_j,t_i,t_j)=\begin{cases}
    s^*=\frac{a_i}{b_i}\frac{1}{4T}+\frac{1}{2}+\frac{(W_j-W_i)}{4T}+\frac{(\Delta_j-\Delta_i)}{4T} & if \ \frac{a_i}{b_i}>\mathbb{H}\\
    s^*=\frac{a_i}{b_i}\frac{(1+\eta)}{4T}+\frac{1}{2}+\frac{(W_j-W_i)}{4T}+\frac{(\Delta_j-\Delta_i)}{4T} & if \ \frac{a_i}{b_i}<\frac{\mathbb{H}}{1+\eta}\\
   s^*= -\frac{\Delta_i}{T} & if \ \frac{\mathbb{H}}{1+\eta}\leq \frac{a_i}{b_i}\leq\mathbb{H}
    \end{cases}
\end{align*}
\end{prop}
The proof, shown in Appendix~\ref{APP:C}, relies on the fact that $U_i^D(y_i,y_j,t_i,t_j)$ is a piecewise function consisting of two concave functions.

The thick red curve in Figure~\ref{fig:example_dg_utility_sharing} panel (b) depicts an example of the optimal taking choice $s^*$ corresponding to $U_i^D(y_i,y_j,t_i,t_j)$ when $\Delta_j=-5$ and $\Delta_i$ varies from $-10$ to $10$. Specifically, when $t_i=\Delta_i+sT$ is negative, the trend gain-loss component of the model bites and the consequent disutility induces the dictator to keep more for him/herself by choosing a higher $s^*$. As $t_i=\Delta_i+sT$ becomes positive, inequality aversion in trends $\sigma_i(t_i,t_j)$ decreases $s^*$. The dashed black curve represents the BO prediction, which is not affected by variations in trends.

As outlined earlier, the functional form in (2) allows us to express
individual heterogeneity through an individual type, which we take as the ratio $a_i/b_i$.
The purely selfish individual is characterized by the limiting case $a_i/b_i \to \infty$,
which implies $s^*=1$.
At the other extreme, the purely altruistic individual is represented by $a_i/b_i=0$.
However, the value of  $s^*$ for the altruistic individual depends on relative incomes and trends. Since those two relative terms can be positive or negative, $s^*$ can deviate from an equal share. A similar argument applies to any type of non fully selfish individual. The next proposition characterizes corner and interiors solutions in terms of individual types by providing an upper bound $\mathbb{U}$ and a lower bound $\mathbb{L}$.

\begin{prop}
  
Define  $\mathbb{U}\equiv W_i-W_j+\Delta_i-\Delta_j+2T$ and $\mathbb{L}\equiv (W_i-W_j+\Delta_i-\Delta_j-2T)(1+\eta)^{-1}$. Then,
  \begin{align*}
      \argmax U_i(u_i,\mu_i,\sigma_i, \tau_i)= \begin{cases}
      s^*=1 & if \ \frac{a_i}{b_i}\geq \mathbb{U}\\
      s^*=0 & if \ \frac{a_i}{b_i}\leq \mathbb{L}\\
      s^*\in (0,1) & if \ \mathbb{L}<\frac{a_i}{b_i}<\mathbb{U}
      \end{cases}
  \end{align*}

  \end{prop} 

In addition to income inequality aversion, two behavioral phenomena coexists in our model: loss aversion in one's own trend and inequality aversion in relative trends. The next two propositions disentangle the two separate effects.

\begin{prop}
Suppose $\Delta_i=\Delta_j$. Then  $s^*$ (weakly) increases in $[0,1]$ when $\Delta_i$ decreases.
\end{prop}

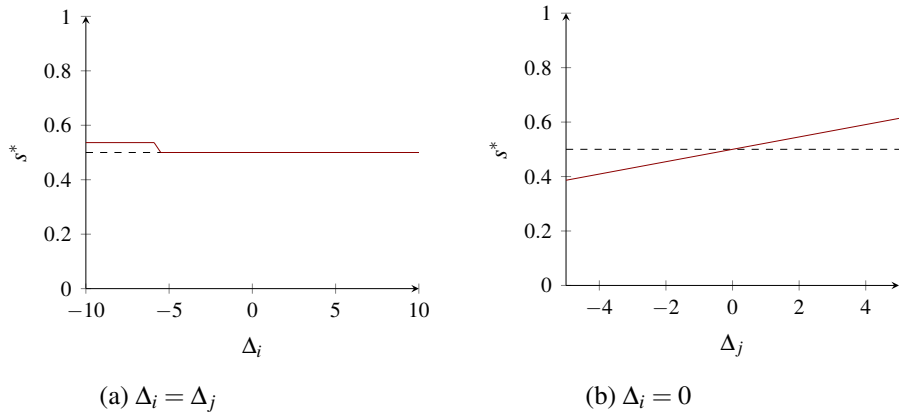
\begin{figure}[h]
\hspace{-2cm}
\centering
    \begin{minipage}[t]{.3\textwidth}
     \begin{tikzpicture}[thick, scale=0.9]
\begin{axis}[
    axis lines = left,
    xlabel = $\Delta_i$,
    ylabel = {$s^*$},
    ymin=0,
    ymax=1
]

\addplot [dashed,
    domain=-10:10, 
    samples=100, 
    color=black,
]
{0.5};

\addplot [
    domain=-5.5:10, 
    samples=100, 
    color=darkred,
]
{0.5};

\addplot [
    domain=-5.9:-5.5, 
    samples=100, 
    color=darkred,
]
{-x/11};

\addplot [
    domain=-10:-5.9, 
    samples=100, 
    color=darkred,
]
{(1.8)/22+0.5-(1/22)};

\end{axis}
\end{tikzpicture}
      \subcaption{ $\Delta_i=\Delta_j$}
    \end{minipage}
    \hspace{2cm}
    \begin{minipage}[t]{.3\textwidth}
       \begin{tikzpicture}[thick, scale=0.9]
\begin{axis}[
    axis lines = left,
    xlabel = $\Delta_j$,
    ylabel = {$s^*$},
       ymin=0,
    ymax=1
]

\addplot [dashed,
    domain=-5:5, 
    samples=100, 
    color=black,
]
{0.5};

\addplot [
    domain=-5:5, 
    samples=100, 
    color=darkred,
]
{0.5+x/44};

\end{axis}
\end{tikzpicture}
        \subcaption{ $\Delta_i=0$}
    \end{minipage}  
    \caption{An example of $s^*$  for $a_i/b_i=2$ $W_i=12$, $W_j=10$, $T=11$ and 
$\eta=0.8$. The thick line indicates our model predictions and the dashed line indicates BO predictions.}
 \label{fig:example_propositions2_3}

\end{figure}

The thick red curve in Figure~\ref{fig:example_propositions2_3} panel (a) illustrates Proposition $3$ with an example of how $s^*$ varies when we change an individual's trend while keeping other factors constant.
There is a region where the trend ($t_i=\Delta_i+sT$) becomes negative and $s^*$ increases. As the trend becomes positive, the model reduces to the BO-type model, which is displayed by the dashed black curve.\footnote{Recall that the trend being positive means that $t_i>0$ (not  $\Delta_i>0$) because the choice of $s$ affects the trend.} We show in the next proposition that, given any individual trend, a lower relative trend decreases the dictator's sharing, i.e., $s^*$ increases when $\Delta_j$ increases.

\begin{prop}Suppose $\Delta_i$ is constant. Then 
$s^*$ (weakly) increases in $[0,1]$ when $\Delta_j$ increases.
\end{prop}

The thick red curve in Figure~\ref{fig:example_propositions2_3} panel (b) illustrates Proposition $4$ with an example of $s^*$ when we vary the other individual's trend, keeping everything else constant. In this case, our prediction reduces to the BO prediction when the relative trend is zero. The difference is that the dictator shares less (more) to compensate a negative (positive) relative trend.

\section{Hypotheses}
\label{sec:hypo}

We posit that individuals with decreasing incomes experience disutility and take from others to be compensated. Specifically, we assume two forms of disutility: (i) intra-personal disutility created by a negative trend, and (ii) inter-personal disutility created by a negative relative trend. We formalized this idea in Section~\ref{sec:model}. In this section, we provide hypotheses for each role in the experiment: High and Low Earners. Keep in mind that, before redistribution takes places, High Earners face the same sum of incomes and income inequality over a period in every treatment, and that the same is true for Low Earners.

First, High Earners experience a negative trend \textsc{IntraDecrease} and a negative relative trend in \textsc{CatchingUp}. In contrast, they do not experience a negative trend or negative relative trend \textsc{Stable}, \textsc{IntraIncrease}, and \textsc{IntraInterChange}. In line with Propositions 3 and 4, we predict the following, which does not distinguish between absolute and relative negative trends.

\begin{Hypothesis}
 High Earners give less in \textsc{IntraDecrease} and  \textsc{CatchingUp} \ than in \ \textsc{Stable}, \textsc{IntraIncrease},  and \ \textsc{IntraInterChange}.
\end{Hypothesis}


Second, Low Earners encounter a negative trend in \textsc{IntraDecrease} and a negative absolute and relative trend in \textsc{IntraInterChange}. In contrast, they do not face a negative trend in \textsc{Stable}, \textsc{IntraIncrease}, and \textsc{CatchingUp}. In line with Propositions 3 and 4, we make the following prediction.

\begin{Hypothesis}
 Low Earners give less in \textsc{IntraInterChange} and \textsc{IntraDecrease} than in \textsc{Stable}, \textsc{IntraIncrease} \, and \textsc{CatchingUp}.
\end{Hypothesis}

Low Earners face an absolute and relative negative trend only in \textsc{IntraInterChange}, and at most a decreasing trend in the other treatments. Therefore, consistent with Proposition 4, we also predict the following.

\begin{Hypothesis}
 Low Earners give less in \textsc{IntraInterChange} than in \textsc{IntraDecrease}, \textsc{Stable}, \textsc{IntraIncrease}, and \textsc{CatchingUp}.
\end{Hypothesis}

Similarly, since Low Earners only experience a negative trend in \textsc{IntraDecrease}, we can make the following prediction, which is consistent with Proposition 4.

\begin{Hypothesis}
Low Earners give less in \textsc{IntraInterChange} than in \textsc{IntraDecrease}.
\end{Hypothesis}




\section{Estimating the Effect of Decreasing Incomes}
\label{sec:results}

We first provide summary statistics regarding the amounts shared in the different treatments. Then we proceed to test the effect of decreasing wages within a period on giving at the end of that period. We make a distinction between the full sample and participants with sufficiently-strong social preferences, whom our framework predicts to be the only ones to become less generous when experiencing decreasing wages (see Proposition 2 in Section~\ref{sec:model}).

\subsection*{Summary Statistics}
\label{subsec:results_sumstats}

Table~\ref{tab:sum} presents the average amount from the 11-Euro joint account that participants give to the other participant. We provide the data for all participants and then for those participants with sufficiently-strong social preferences, i.e., who give a minimum amount in the baseline treatment with constant wages (\textsc{Stable}). We use a minimum amount of 2 Euro, but the results are robust to using different cutoff points.\footnote{We chose a cutoff point that denotes a sufficiently-strong generosity in the absence of absolute or relative wage changes, but provide tests of our hypotheses using different cutoffs to show that this specific cutoff does not drive the results. Patterns in descriptive statistics are similar if we instead take cutoffs of 3, 1 and even 0.25 Euro (see Table~\ref{tab:sum_othercutoffs}). Cutoffs higher than 2 Euro yield samples that are smaller and less suitable for separate analyses of the behavior of High Earners and Low Earners (e.g., 81 participants for 3 Euro (44 High Earners and 37 Low Earners) and 69 participants for 4 Euro (38 High Earners and 31 Low Earners)).} Overall, mean giving is 1.53 Euro (SD = 2.23 Euro; \textit{N} = 1,470 observations, 294 participants) or 14\% of the joint account for the entire sample, and 4.12 Euro (SD = 1.93 Euro; \textit{N} = 455 observations, 91 participants) or 37\% of the joint account for those with social preferences. High Earners and Low Earners exhibit similar mean giving (for all participants, High Earners give 1.57 Euro (SD = 2.13 Euro; \textit{N} = 147) and Low Earners give 1.49 (SD = 2.02; \textit{N} = 147); \textit{t}-test and Mann-Whitney-Wilcoxon rank test \textit{p}-values $\geq$ 0.599; the difference is smaller and \textit{p}-values larger for socially-inclined participants). The giving patterns are more pronounced for participants with social preferences: High Earners are less generous in \textsc{CatchingUp} and perhaps in \textsc{IntraDecrease}; Low Earners are less generous in \textsc{IntraInterChange}.

Compared to dictator games in general, average giving in our experiment is on the lower side of the spectrum found in the literature (see meta-study by \citet{engel2011dictator} and comment by \citet{zhang2014effects}; average giving in the dictator game is $28.3\%$). Factors present in our study can reduce generosity, namely using a student sample, endowing recipients, repeating the game, dictators earning the money that they can redistribute, and the option of taking money from others, although having deserving recipients who earned the money that the dictator redistributes can increase giving. Moreover, role uncertainty about who will give and who will receive has also been shown to increase pro-sociality in dictator games \citep{iriberri2011role}. 

\begin{table}[t!]	
\centering		
\adjustbox{max width=\textwidth}{

    \begin{threeparttable}
\caption{Summary statistics}\label{tab:sum}
\begin{tabular}{rccccc}								
\toprule								
\toprule	
 &\multicolumn{5}{c}{Amount given}\\ \cline{2-6}
  &\multicolumn{2}{c}{All participants} & &\multicolumn{2}{c}{Participants with social preferences}\\ \cline{2-3} \cline{5-6}

 Treatment   & High Earners  & Low Earners & & High Earners  &Low Earners \\
 &	Mean (SD)	&	Mean 	(SD)  & &	Mean (SD)	&	Mean 	(SD)	\\
\midrule							
\textsc{Stable}	& 1.62		& 1.49	& & 4.74 &	4.54	\\
	& (2.35)		&	(2.18) & & (1.46) & (1.60)	\\
[.2em]
\textsc{IntraDecrease}	&1.57 		& 1.48	& & 3.86 & 4.14		\\
	&	(2.42)		&		(2.15)	& & (2.17) &  (1.99)	\\
[.2em]
\textsc{CatchingUp}	& 1.41	& 1.58	& &  3.61 & 4.19			\\
(High Earners inter-personal decrease)	&	(2.20)	&		(2.15)	& & (1.98) & (1.92)\\
[.2em]
\textsc{IntraInterChange}	& 1.62	& 1.42	& & 4.16 &	 3.71	\\
(Low Earners inter- and intra-personal decrease)	&	(2.34)		&		(2.03)	& & (2.01) & (1.99)	\\
[.2em]
\textsc{IntraIncrease}	&1.63	&1.45	 & & 4.29    & 3.99	\\    
	&	(2.39)		&		(2.13) & & (1.90) & (2.09)	\\
\bottomrule								
\(N\)       &  147  & 147 & & 48   &  43    \\
[.2em] \hline
\bottomrule								
\bottomrule								
\end{tabular}	

\begin{tablenotes}
\item \footnotesize Note:  Participants could give any amount between 0 and 11 Euro from the 11-Euro joint account. The remaining amount was credited to their own account. Participants with social preferences are those exhibiting sufficiently-strong advantageous inequality aversion, i.e., they give at least 2 Euro in the baseline treatment \textsc{Stable}. Table~\ref{tab:sum_othercutoffs} shows that other cutoff points such as at least 3 Euro, 1 Euro, and 0.25 Euro in \textsc{Stable} provide similar pictures.
\end{tablenotes}
\end{threeparttable}
}
\end{table}		

\begin{figure}[htbp!]
\centering 
\begin{subfigure}[t]{6.8cm}
\includegraphics[width=6.8cm]{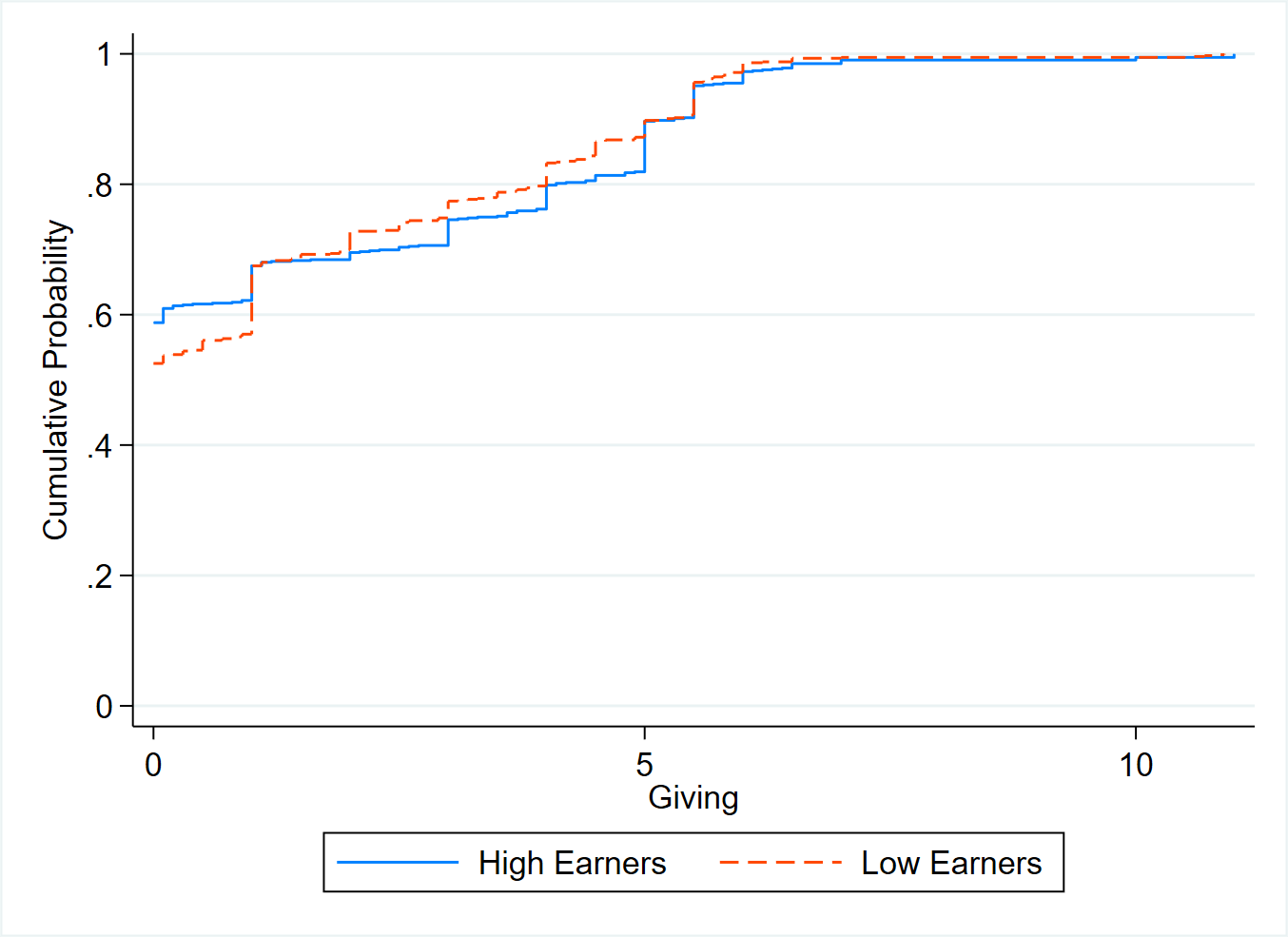}
 \caption{All participants}
\end{subfigure}
\begin{subfigure}[t]{6.8cm}
\includegraphics[width=6.8cm]{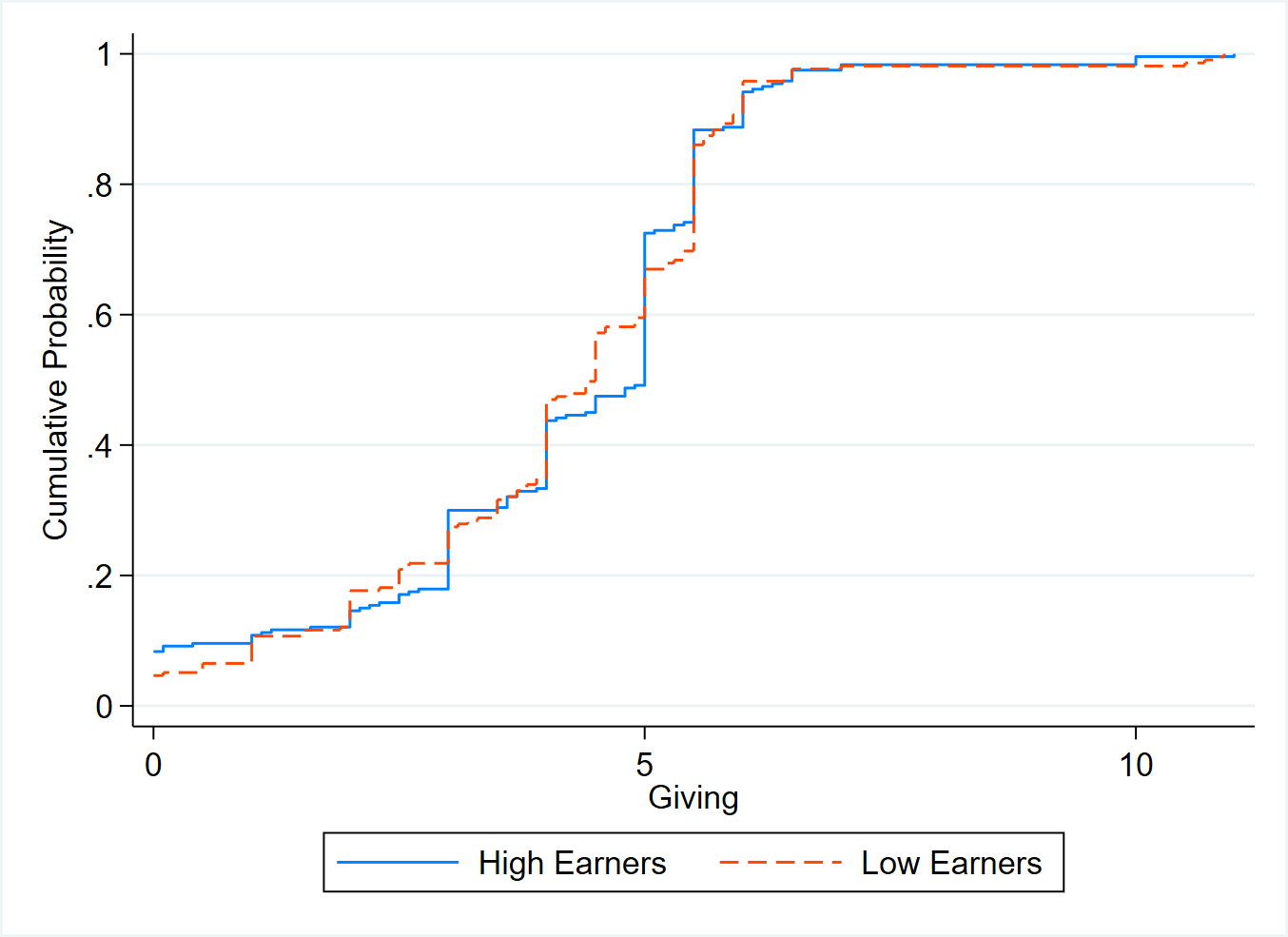}
 \caption{Participants with social preferences}
\end{subfigure}
 \caption{Cumulative distribution functions of giving}\label{fig:CDF}
 
\end{figure}

Figure~\ref{fig:CDF} shows the cumulative distribution of the amounts given, for High Earners and Low Earners, separately for the entire sample and for those participants with social preferences. There is extensive lower-bound censoring in the full sample---participants give nothing in 56\% of decisions---and this censoring mostly vanishes for socially-inclined participants.

\subsection*{Tests}
\label{subsec:results_tests}

Table~\ref{tab:results} shows the estimated effect of the wage decrease treatments on giving. We use Tobit Random Effects regressions (Tobit RE) for the heavily-censored full sample, and Random Effects regressions (RE) for participants with sufficiently-strong social preferences. In Appendix~\ref{APP:D}, Table~\ref{tab:results_fe} shows that results are very similar with Fixed Effects (FE) regressions for participants with sufficiently-strong social preferences, and Table~\ref{tab:results_socialpreferences300}, ~\ref{tab:results_socialpreferences100}, and ~\ref{tab:results_socialpreferences025} show that the RE results are also very similar if we use alternative cutoff points for defining participants with social preferences (giving at least 3 Euro, 1 Euro, and 0.25 Euro in \textsc{Stable}). Any decrease includes all treatments with decreasing incomes (\textsc{IntraDecrease} and \textsc{CatchingUp} for Higher Earners and \textsc{IntraDecrease} and \textsc{IntraInterDecrease} for Low Earners). Any strictly intra decrease denotes all treatments with only intra-personal decreasing incomes (\textsc{CatchingUp} for High Earners and \textsc{IntraInterDecrease} for Low Earners), and Any inter decrease denotes all treatments with inter-personal decreasing incomes (\textsc{CatchingUp} for High Earners and \textsc{IntraInterDecrease} for Low Earners).\footnote{\textsc{IntraInterDecrease} combines intra- and inter-personal decreases for Low Earners.} Columns 1--6 display the results for all participants, and columns 7--20 for participants with sufficiently-strong social preferences. The results are presented for All Earners or separately for High Earners and Low Earners. We include period dummies, and a dummy for High Earners when we employ all Earners in the regression. 

Importantly, for socially-inclined participants, we present the results including and excluding the \textsc{Stable} treatment. This matters because including \textsc{Stable} in the regression could create a spurious negative relationship between the decreasing treatments and \textsc{Stable}. For instance, suppose that participants' differences in giving choices across treatments simply consist of noise: they randomly give 0 or 2 Euro for every decision. Then, if we compare the other treatments to \textsc{Stable} only for participants with social preferences (defined as those who give at least 2 Euro in \textsc{Stable} where there is no wage trend), those participants will mechanically appear less generous in the other treatments. Reassuringly, our results show that, while excluding \textsc{Stable} qualitatively affects the size of the estimates, it has minor impacts on our conclusions.

To complement this analysis, Table~\ref{tab:results_nonpara} presents Wilcoxon signed-rank tests roughly equivalent to several tests in Table~\ref{tab:results}. We write roughly because those tests only provide pairwise comparisons of treatments. Although they reduce statistical power, their advantage is to relax the normality of error terms assumption. The conclusions that we can draw from those tests are similar to those using regressions, but the results are not always significant. We indicate in footnotes whenever differences arise.

We make three sets of observations regarding the analyses reported in Table~\ref{tab:results}. First, when considering High and Low Earners together (columns 1-2 and 7-8), any decrease in income reduces giving at marginally significant levels for the full sample and at highly significant levels for participants with social preferences.\footnote{The effect is significant for some pairwise comparisons using rank-based tests on the full sample in Table~\ref{tab:results_nonpara}. If we separate Earners by their type, the effect of any decrease is not significant for High and Low Earners alone in the full sample (columns 3-5). For socially-inclined participants (columns 11-12 and 15-16), the effect is still highly significant for High Earners, and not significant for Low Earners in our preferred robust specification without \textsc{Stable}.} For participants with social preferences, the effect amounts to approximately $-0.5$ Euro; relative to their mean giving of 4.12 (SD = 1.93) Euro, this corresponds to 1/4 of a standard deviation. Once we separate any strictly intra-personal decreases and any inter-personal decreases (column 2 and 9-10), we observe that only inter-personal decreases have a significant impact in the full sample and in our robust specification excluding the \textsc{Stable} treatment for participants with social preferences (the effect of any strictly intra-personal decreases is significant under our basic specification). The Wald tests indicate that difference between any strictly intra-personal decreases and any inter-personal decreases is not significant in the full sample, and marginally significant under our preferred robust specification for social participants (just below the 0.05 significance threshold under our basic specification). We provide a first result of the effect of decreasing incomes among all Earners because we believe it informative to provide an aggregate picture, but this first result does not correspond to any specific hypothesis. We write it as follows.

\begin{res}
 Among all Earners with social preferences, decreasing incomes reduce average giving compared to non-decreasing incomes; inter-personal decreases have a significant effect, which is marginally stronger than that of strictly intra-personal decreases. Decreasing incomes have a marginally significant impact in the full sample.
\end{res}

\begin{landscape}

\begin{table}[h!]							
\centering	
\adjustbox{max width=20cm}{
\begin{threeparttable}
\caption{Regression tests of the effect of decreasing wages on giving }\label{tab:results}

\begin{tabular}{{r}c*{24}{c}}
\hline\hline
            &\multicolumn{1}{c}{(1)}&\multicolumn{1}{c}{(2)}&&\multicolumn{1}{c}{(3)}&\multicolumn{1}{c}{(4)}&&\multicolumn{1}{c}{(5)}&\multicolumn{1}{c}{(6)}&&\multicolumn{1}{c}{(7)}&\multicolumn{1}{c}{(8)}&\multicolumn{1}{c}{(9)}&\multicolumn{1}{c}{(10)}&&\multicolumn{1}{c}{(11)}&\multicolumn{1}{c}{(12)}&\multicolumn{1}{c}{(13)}&\multicolumn{1}{c}{(14)}&&\multicolumn{1}{c}{(15)}&\multicolumn{1}{c}{(16)}&\multicolumn{1}{c}{(17)}&\multicolumn{1}{c}{(18)}&\multicolumn{1}{c}{(19)}&\multicolumn{1}{c}{(20)}\\
            & Tobit RE & Tobit RE && Tobit RE & Tobit RE && Tobit RE & Tobit RE && RE & RE & RE & RE && RE & RE & RE & RE && RE & RE & RE & RE & RE & RE \\
[.5em]
            & \multicolumn{8}{c}{All Participants} && \multicolumn{16}{c}{Participants with Social Preferences} \\ \cline{2-9} \cline{11-26}
 & \multicolumn{2}{c}{All Earners} && \multicolumn{2}{c}{High Earners} && \multicolumn{2}{c}{Low Earners} && \multicolumn{4}{c}{All Earners} && \multicolumn{4}{c}{High Earners} && \multicolumn{6}{c}{Low Earners} \\ \cline{2-3} \cline{5-6} \cline{8-9} \cline{11-14} \cline{16-19} \cline{21-26}

& & && & && & && & & & && & & & && & & & & \\
Any decrease    &      $-$0.176$^{*}$ &   &&      $-$0.304   &               &&     $-$0.104   &               &&      $-$0.483$^{***}$ &      $-$0.331$^{***}$ &  &  &&     $-$0.636$^{***}$ &       $-$0.507$^{**}$ &               &               &&      $-$0.322$^{***}$ &      $-$0.151   &               &               &               &                \\
            &     (0.105) &    &&    (0.187)   &               &&    (0.114)   &               &&     (0.117)   &     (0.122)  &  &    &&     (0.200)   &      (0.210)   &               &               &&    (0.115)  &     (0.120)   &               &               &               &                \\
[1em]
Any strictly intra decrease & & $-$0.055 && & && & && & & $-$0.301$^{**}$ & $-$0.159 & && & & & && & & & & \\
& & (0.133) && & && & && & & (0.118) & (0.134) && & & & && & & & & \\
[1em]
Any inter decrease & & $-$0.295$^{**}$ && & && & && & & $-$0.660$^{***}$ & $-$0.502$^{***}$ && & & & && & & & & \\
& & (0.133) && & && & && & & (0.172) & (0.166) && & & & && & & & & \\
& & && & && & && & & & && & & & && & & & & \\
\hline
Wald test \textit{p}-value & & 0.142 && & && & && & & 0.050 & 0.054 && & & & && & & & & & \\
\hline
& & && & && & && & & & && & & & && & & & & \\
\textsc{IntraDecrease}      &        &         &&             &      $-$0.170   &&              &      0.012   &&              &       &     &          &&              &                &      $-$0.492$^{***}$ &      $-$0.367$^{*}$  &&              &               &     $-$0.094   &      0.066   &               &                \\
            &      &           &&              &     (0.235)   &&              &     (0.143)   &&              &      &    &           &&              &                &     (0.182)   &     (0.203)   &&              &               &      (0.135)   &     (0.167)   &               &                \\
[1em]
\textsc{CatchingUp}      &      &           &&              &      $-$0.437$^{*}$  &&    &              &&               &      &   &          &&               &                &      $-$0.781$^{**}$ &      $-$0.653$^{**}$ &&              &               &               &               &               &                \\
            &       &          &&              &     (0.236)   &&    &              &&               &     &   &            &&              &                &     (0.308)   &     (0.308)   &&              &               &               &               &               &                \\
[1em]
\textsc{IntraInterDecrease}      &     &            &&              &               &&              &      $-$0.219   &&              &      &    &           &&              &                &               &               &&              &               &      $-$0.551$^{***}$ &      $-$0.376$^{**}$ &      $-$0.527$^{***}$ &      $-$0.398$^{**}$  \\
            &       &          &&              &               &&              &     (0.144)   &&              &       &     &          &&              &                &               &               &&              &               &     (0.181)    &     (0.164)    &     (0.182)   &     (0.168)    \\
[1em]
\hline
Wald test \textit{p}-value & & && & 0.355 && & 0.188 && & & & && & & 0.352 & 0.349 && & & 0.038 & 0.048 & & \\
\hline
Constant      &      $-$0.446 &  $-$0.441$^{***}$  &&     $-$0.902$^{*}$  &      $-$0.909$^{*}$  &&      0.194   &       0.208   &&      4.484$^{***}$ &       4.199$^{***}$ & 4.491$^{***}$ & 4.208$^{***}$ &&      4.556$^{***}$ &        4.243$^{***}$ &       4.547$^{***}$ &       4.231$^{***}$ &&      4.361$^{***}$ &       4.133$^{***}$ &       4.398$^{***}$ &       4.187$^{***}$ &       4.378$^{***}$ &       4.209$^{***}$  \\
            &     (0.426) &  (0.425)  &&    (0.520)   &     (0.519)   &&    (0.399)   &     (0.399)  &&     (0.202)   &     (0.248) & (0.201) &  (0.246)  &&    (0.213)   &      (0.272)    &     (0.215)   &     (0.276)   &&   (0.247) & (0.271)   &     (0.246)    &     (0.268)   &     (0.248)   &     (0.271)     \\
\hline
Without \textsc{Stable} & & && & && & && & \checkmark  &  & \checkmark && & \checkmark & & \checkmark && & \checkmark & & \checkmark & & \checkmark \\
Period dummies & \checkmark & \checkmark && \checkmark & \checkmark && \checkmark & \checkmark && \checkmark & \checkmark & \checkmark & \checkmark && \checkmark & \checkmark & \checkmark & \checkmark && \checkmark & \checkmark & \checkmark & \checkmark & \checkmark & \checkmark \\
High Earners dummy & \checkmark & \checkmark && & && & && \checkmark & \checkmark & \checkmark & \checkmark && & & & && & & & & & \\

\hline
\(N\)       &        1470    &  1470  &&        735   &         735   &&        735   &         735   &&        455   &         364   &  455   &  364   &&        240   &          192   &         240   &         192   &&        215   &         172   &         215   &         172   &         215   &         172    \\
\hline\hline
\end{tabular}

\begin{tablenotes}
\item \footnotesize Note: Giving is indicated in Euro. Participants with social preferences are those exhibiting sufficiently-strong advantageous inequality aversion, i.e., they give at least 2 Euro in the baseline treatment \textsc{Stable}. Table~\ref{tab:results_socialpreferences300}, ~\ref{tab:results_socialpreferences100}, and ~\ref{tab:results_socialpreferences025} show that the results are very similar if we use alternative cutoff points for defining participants with social preferences (giving at least 3 Euro, 1 Euro, and 0.25 Euro in \textsc{Stable}). Any decrease denotes \textsc{IntraDecrease} and \textsc{CatchingUp} for High Earners, and \textsc{IntraDecrease} and \textsc{IntraInterDecrease} for Low Earners. Any strictly intra decrease denotes \textsc{IntraDecrease} for High and Low Earners. Any inter decrease denotes \textsc{CatchingUp} for High Earners and \textsc{IntraInterDecrease} for Low Earners. All treatments are included in the regression, except when we indicate that we remove the \textsc{Stable} treatment. We indicate \textit{p}-values from Wald test on the equality of the coefficients of Any strictly intra decrease and Any inter decrease, and on the equality of \textsc{IntraInterDecrease} and \textsc{IntraDecrease} or \textsc{CatchingUp}. The coefficient of the High Earners dummy is always far from the 10\% significance threshold. Table~\ref{tab:results_fe} also shows that results are very similar with Fixed Effects regressions for participants with sufficiently-strong social preferences. Standard errors are in parentheses. $^{*}$ \textit{p}$<$0.10, $^{**}$ \textit{p}$<$0.05, $^{***}$ \textit{p}$<$0.01.
\end{tablenotes}
\end{threeparttable}
}	
\end{table}

\begin{table}[h!]							
\centering	
\adjustbox{max width=16cm}{
\begin{threeparttable}
\caption{Wilcoxon signed-rank tests of the effect of decreasing wages on giving}\label{tab:results_nonpara}

\begin{tabular}{{r}c*{11}{c}}
\hline\hline
&  & \multicolumn{1}{c}{(1)} & \multicolumn{1}{c}{(3)} & \multicolumn{1}{c}{(4)} & \multicolumn{1}{c}{(5)} & & \multicolumn{1}{c}{(6)} & \multicolumn{1}{c}{(7)} & \multicolumn{1}{c}{(8)} & \multicolumn{1}{c}{(9)} \\
& & \multicolumn{4}{c}{All Participants} & & \multicolumn{4}{c}{Social Participants} \\ \cline{3-6} \cline{8-11}
& & \textsc{Stable} & \textsc{IntraIncrease} & \textsc{OtherNoDecrease} & \textsc{IntraDecrease} & & \textsc{Stable} & \textsc{IntraIncrease} &  \textsc{OtherNoDecrease} & \textsc{IntraDecrease} \\
\hline
All Earners & \textsc{IntraDecrease} & 0.042 & 0.268 & 0.050 &  & & $<$0.001 & 0.643 & 0.245 & \\
& \textsc{OtherDecrease} & 0.058 & 0.147 & 0.026 &  & & $<$0.001 & 0.014 & 0.008 & \\
& \textit{N} & 294 & 294 & 294 &  & & 91 & 91 & 91 & \\
[1em]
High Earners & \textsc{IntraDecrease} & 0.145 & 0.172 &  0.198 &  & & 0.001 & 0.216 & 0.268 & \\
& \textsc{CatchingUp} & 0.045 & 0.064 & 0.053 & 0.455 & &  $<$0.001 & 0.032 & 0.087 & 0.309 \\
& \textit{N} & 147 & 147 & 147 & 147 & & 48 & 48 & 48 & 48 \\
[1em]
Low Earners & \textsc{IntraDecrease} & 0.173 & 0.813 & 0.140 &  & & 0.004 & 0.536 & 0.618 & \\ 
& \textsc{IntraInterDecrease} & 0.439 & 0.727 & 0.202 & 0.968 & & 0.001 & 0.176 & 0.037 & 0.241 \\
& \textit{N} & 147 & 147 & 147 & 147 & & 43 & 43 & 43 & 43 \\
\hline
\hline\hline
\end{tabular}

\begin{tablenotes}
\item \footnotesize Note: \textsc{OtherDecrease} denotes \textsc{CatchingUp} for High Earners and \textsc{IntraInterDecrease} for Low Earners;  \textsc{OtherNoDecrease} denotes \textsc{IntraInterDecrease} for High Earners and \textsc{CatchingUp} for Low Earners.
\end{tablenotes}
\end{threeparttable}
}	
\end{table}	

\end{landscape}

Second, among socially-inclined High Earners (columns 11--14), the intra-personal (\textsc{IntraDecrease}) and the inter-personal (\textsc{CatchingUp}) decreases in income reduce average giving, at marginal and significant levels, respectively, under our preferred robust specification without \textsc{Stable}.\footnote{Both effects are significant under our other specification with \textsc{Stable}. If we use the non-parametric tests of Table~\ref{tab:results_nonpara}, only \textsc{CatchingUp} is significant or marginally significant in pairwise comparisons with the two non-decreasing income treatments other than \textsc{Stable}.} Although the effect of inter-personal decrease (approx. $-0.7$ Euro) is qualitatively larger than that of the intra-personal decrease (approx. $-0.4$ Euro), the Wald tests show that the two are statistically indistinguishable. Furthermore, only the effect of inter-personal decrease is marginally significant with the full sample (column 4). The evidence is in line with Hypothesis 1 for participants with social preferences, but not for the full sample. We therefore write our second result as follows.

\begin{res}
Among High Earners with social preferences, decreasing incomes reduce average giving compared to non-decreasing incomes; inter- (\textsc{CatchingUp}) and intra-personal (\textsc{IntraDecrease}) decreases have significant and marginally significant effects, respectively, but the two are statistically indistinguishable. Only the effect of inter-personal decreases is marginally detectable in the full sample.
\end{res}

Third, among Low Earners with social preferences (columns 15--18), the joint intra- and inter-personal decrease in incomes (\textsc{IntraInterDecrease}) significantly reduces average giving (approx. $-0.4$ Euro), but the point estimate of intra-personal decrease (\textsc{IntraDecrease}) is nearing 0 and not significant. The table's Wald test shows that the effect of the joint intra- and inter-personal decrease is significantly larger than the the intra-personal effect.\footnote{If we use the non-parametric tests of Table~\ref{tab:results_nonpara}, only \textsc{IntraInterDecrease} is significant for Low Earners in one of the pairwise comparisons with the two non-decreasing income treatments other than \textsc{Stable}. Furthermore, \textsc{IntraInterDecrease} is not significantly larger than \textsc{IntraDecrease}. Overall, we primarily rely on the RE linear regressions for those tests because (i) a within-subject approach is statistically much more powerful \citep{bellemare2014statistical}, (ii) although not strictly normally distributed, giving and regression residuals for participants with social preferences do not visually look too far off (e.g., see Figure~\ref{fig:CDF} (b) for CDF of giving), and (3) as we show in the next subsection, order effects do not appear to influence our findings.} The joint intra- and inter-personal decrease is also significantly larger than all other schemes bundled together (columns 19-20). Moreover, similarly as for High Earners, no effect of any of the two types of income decreases has an effect on Low Earners in the full sample. Overall, for Low Earners with social preferences, the evidence goes against Hypothesis 2, which ascertains that both types of decreases would reduce giving. However, the evidence falls in line with Hypotheses 3 and 4, which predict that the joint intra- and inter-personal decrease would lower giving and that its effect would be larger than that of the intra-personal decrease. The full sample supports none of those three hypotheses. We thus write our third result as follows.

\begin{res}
Among Low Earners with social preferences, the joint intra- and inter-personal income decrease (\textsc{IntraInterDecrease}) significantly reduces average giving compared to non-decreasing incomes, and the intra-personal decrease (\textsc{IntraDecrease}) does not; the effect of the first is significantly larger than that of the second. No effect is detectable in the full sample.
\end{res}

To sum up, although income inequality over a period is the same in each treatment, we find strong evidence of a negative effect of decreasing incomes within a period on giving among participants with social preferences.  This is the case for all Earners joined together, and, when we consider the different roles in the experiment separately, for High and Low Earners evaluated separately. Overall, the evidence is strong for inter-personal decreases rendering participants more selfish and weak for intra-personal decreases. However, the difference between the two is itself limited: the inter-personal decrease is stronger than the intra-personal decrease at marginal significance levels for all Earners evaluated together, and right below significance levels for Low Earners.\footnote{More precisely, the effect of any inter decrease is stronger than that of any strictly intra decrease.} For the sample as a whole, there is limited (marginally significant) evidence for the negative effect of decreasing incomes on giving.

\subsection*{Evidence that an Increasing Relative Income Trend does not Increase Sharing}
\label{subsec:results_increase}

\begin{table}[h!]							
\centering	
\adjustbox{max width=\textwidth}{
\begin{threeparttable}
\caption{Tests of the effect of increasing relative wages on giving for High Earners, excluding treatments with decreasing wages}\label{tab:results_relativeincrease}

\begin{tabular}{{r}c*{3}{c}}
\hline\hline
            & \multicolumn{1}{c}{(1)} && \multicolumn{1}{c}{(2)}&  \multicolumn{1}{c}{(3)}\\
            & RE && RE & RE \\
[.5em]
            & \multicolumn{1}{c}{All Participants} & & \multicolumn{2}{c}{Participants with Social Preferences} \\ \cline{2-2} \cline{4-5}
 & \multicolumn{1}{c}{High Earners} && \multicolumn{2}{c}{High Earners} \\ \cline{2-2} \cline{4-5}
&  && &   \\
\textsc{IntraInterDecrease}    &   $-0.044$ && $-0.365^{*}$ & $-0.150$ \\
             &   (0.108)   && (0.191) & (0.198)  \\
[1em]
\hline
Constant       &   1.920 && 4.828  &  4.550   \\
             &   (0.216)   && (0.241) & (0.323)  \\
\hline
Baseline is \textsc{IntraDecrease} \& \textsc{Stable} & \checkmark && \checkmark &  \\
Baseline is \textsc{IntraDecrease} alone & && & \checkmark \\
Period dummies & \checkmark && \checkmark & \checkmark \\
\hline
\(N\)        &   441   && 144 & 96  \\
\hline\hline
\end{tabular}

\begin{tablenotes}
\item \footnotesize Note: Giving is indicated in Euro. The treatments included in the regression are \textsc{IntraInterDecrease} and the baseline treatment(s) indicated at the bottom of the table. Standard errors are in parentheses. $^{*}$ \textit{p}$<$0.10, $^{**}$ \textit{p}$<$0.05, $^{***}$ \textit{p}$<$0.01.
\end{tablenotes}
\end{threeparttable}
}	
\end{table}	

We did not make a hypothesis about how a positive relative income trend might affect generosity. Nevertheless, our formal model differs with our original intuition on this matter. Specifically, through Proposition 4, it predicts, if anything, a positive effect. We evaluate this possibility by using treatment \textsc{IntraInterChange}, where High Earners experience a decreasing relative trend. We compare their sharing behavior in \textsc{IntraInterChange} relative to \textsc{IntraIncrease} and \textsc{Stable}. Table~\ref{tab:results_relativeincrease} shows the analyses, first for all High Earners and second for socially-inclined High Earners, without and then with removing the \textsc{Stable} treatment for robustness. Overall, we find no evidence that a positive relative trend increases sharing. This gives us an additional fourth result, which does not correspond to a hypothesis. 

\begin{res}
Among all High Earners and among High Earners with social preferences, the relative increase in incomes (\textsc{IntraInterDecrease}) does not significantly alter average giving compared to other non-decreasing   incomes.
\end{res}

\subsection*{Evidence that Order Effects Do not Drive our Results}
\label{subsec:results_ordereffects}

\begin{table}[h!]							
\centering	
\adjustbox{max width=\textwidth}{
\begin{threeparttable}
\caption{Tests of the effect of decreasing wages on giving, in first two periods while controlling for possible order effect of starting with a decreasing income treatment}\label{tab:results_firstperiods}

\begin{tabular}{{r}c*{9}{c}}
\hline\hline
            & \multicolumn{1}{c}{(1)} & \multicolumn{1}{c}{(2)} & \multicolumn{1}{c}{(3)} & \multicolumn{1}{c}{(4)} && \multicolumn{1}{c}{(5)}&\multicolumn{1}{c}{(6)}&&\multicolumn{1}{c}{(7)}&\multicolumn{1}{c}{(8)}\\
            & RE & RE & RE & RE && RE & RE && RE & RE \\
[.5em]
            & \multicolumn{10}{c}{Participants with Social Preferences} \\ \cline{2-11}
 & \multicolumn{4}{c}{All Earners} && \multicolumn{2}{c}{High Earners} && \multicolumn{2}{c}{Low Earners} \\
 [.5em]
Period(s) & \multicolumn{4}{c}{\nth{1}-\nth{2}} && \multicolumn{2}{c}{\nth{1}-\nth{2}} && \multicolumn{2}{c}{\nth{1}-\nth{2}} \\ \cline{2-5} \cline{7-8} \cline{10-11}

&  &  &  &  &&  &  &&  & \\
Any decrease    &   $-$0.832$^{***}$ & $-$0.560$^{**}$ &  &  && $-$1.033$^{**}$  & $-$0.692$^{**}$     &&  $-$0.556$^{*}$  & $-$0.456                           \\
             &   (0.275)   &      (0.265) &  &    &&  (0.457)  & (0.284)  && (0.297)  & (0.385)                  \\
[1em]
Any strictly intra decrease &  &  & -0.340 & -0.104 &&  &  &&  & \\
 &  &  & (0.309) & (0.381) &&  &  &&  & \\
[1em]
 Any inter decrease &  &  & $-$1.136$^{***}$ & $-$0.812$^{***}$ &&  &  &&  & \\
 &  &  & (0.367) & (0.306) &&  &  &&  & \\
 &  &  &  &  &&  &  &&  & \\
\hline
Wald test \textit{p}-value  &  &  & 0.075 & 0.104 &&  &  &&  &  \\

\hline
&  &  &  &  &&  &  &&  & \\
Any decrease in \nth{1} period dummy    &  0.324  &    0.250 & 0.293 & 0.197 &&  0.470 & 0.333  &&  0.155 & 0.206 \\
              &     (0.348)   &      (0.432)   & (0.350) & (0.439) && (0.471)  & (0.562)   &&  (0.513)  & (0.629)   \\
[1em]
Constant       &   4.580$^{***}$ &    4.245$^{***}$ &  4.569$^{***}$ & 4.229$^{***}$ && 4.536$^{***}$ & 4.263$^{***}$  && 4.396$^{***}$ & 4.156$^{***}$  \\
             &   (0.203)   &      (0.264) & (0.202) & (0.260)   &&   (0.221)     & (0.284)   &&  (0.313)   & (0.389)     \\
\hline
Without \textsc{Stable} & & \checkmark & & \checkmark &&  & \checkmark &&  & \checkmark \\
Period dummies & \checkmark & \checkmark & \checkmark & \checkmark && \checkmark & \checkmark && \checkmark & \checkmark \\
High Earners dummy & \checkmark & \checkmark & \checkmark & \checkmark && & && & \\

\hline
\(N\)     &     182    &   132  & 182 & 132  &&   96     & 69    &&   86   & 63  \\
\hline\hline
\end{tabular}

\begin{tablenotes}
\item \footnotesize Note: Note:  Giving is indicated in Euro. Any decrease denotes \textsc{IntraDecrease} and \textsc{CatchingUp} for High Earners, and \textsc{IntraDecrease} and \textsc{IntraInterDecrease} for Low Earners. Any strictly intra decrease denotes \textsc{IntraDecrease} for High and Low Earners. Any inter decrease denotes \textsc{CatchingUp} for High Earners and \textsc{IntraInterDecrease} for Low Earners. All treatments are included in the regression, except when we indicate that we remove the \textsc{Stable} treatment. We indicate \textit{p}-values from Wald test on the equality of the coefficients of Any strictly intra decrease and Any inter decrease. Standard errors are in parentheses. $^{*}$ \textit{p}$<$0.10, $^{**}$ \textit{p}$<$0.05, $^{***}$ \textit{p}$<$0.01.
\end{tablenotes}
\end{threeparttable}
}	
\end{table}	

We provide evidence that order effects do not drive our results by re-running the regressions from Table~\ref{tab:results} concerning the effect of any decrease in  income as well as any strictly intra decrease and any inter decrease for participants with social preferences, but this time only using the first two periods and including a dummy for facing a treatment with any  income decrease in the first period. Table~\ref{tab:results_firstperiods} shows the estimates for High and Low Earners together (columns 1--4), High Earners (column 5-6), and Low Earners (column 7-8). Keeping in mind the lower statistical power, the point estimates and significance levels are similar as for the corresponding Table~\ref{tab:results} estimations; the dummy for starting with any decreasing income is never significant.\footnote{We do not present estimation of the effect of individual treatments because the number of observations is much reduced by using only 2 periods.}

\subsection*{Evidence that First-Period Experience Does not Stop Effect of Decreasing Wages}
\label{subsec:results_firstperiod}

We show that having some experience with the experiment, in the form of going through the first period, does not stop participants' reaction to decreasing wages in the next periods. Table~\ref{tab:results_firstperioddummy} reproduces basic regressions from Table~\ref{tab:results} for social participants, but adding a term for the interaction between the first period and Any decrease. We do not distinguish between the two types of decreases to avoid diluting statistical power. The table shows that the negative effect of any decrease is not restricted to the first period as the coefficient of any decrease is significant. Furthermore, the interaction term between the first period and any decrease is negative---suggesting a possible greater impact in the first period---but not significant.

\begin{table}[h!]							
\centering	
\adjustbox{max width=\textwidth}{
\begin{threeparttable}
\caption{Tests of the effect of decreasing wages on giving, adding a dummy for the first encounter with decreasing wages}\label{tab:results_firstperioddummy}

\begin{tabular}{{r}c*{7}{c}}
\hline\hline
            & \multicolumn{1}{c}{(1)} & \multicolumn{1}{c}{(2)} && \multicolumn{1}{c}{(3)}&\multicolumn{1}{c}{(4)}&&\multicolumn{1}{c}{(5)}&\multicolumn{1}{c}{(6)}\\
            & RE & RE && RE & RE && RE & RE \\
[.5em]
            & \multicolumn{8}{c}{Participants with Social Preferences} \\ \cline{2-9}
 & \multicolumn{2}{c}{All Earners} && \multicolumn{2}{c}{High Earners} && \multicolumn{2}{c}{Low Earners} \\ \cline{2-3} \cline{5-6} \cline{8-9}
&  &  &&  &  &&  & \\
Any decrease    &   $-$0.365$^{***}$ & $-$0.259$^{**}$ && $-$0.511$^{***}$  &  $-$0.407$^{**}$    && $-$0.217$^{**}$   & $-$0.103 \\
             &   (0.100)   &   (0.115)   &&  (0.160)  & (0.190)  && (0.103)  & (0.112) \\
[1em]
Any decrease $\times$ \nth{1} period  & $-$0.610 & $-$0.381 && $-$0.745  &  $-$0.553  &&  $-$0.497  & $-$0.258 \\
             &   (0.421)   &   (0.442)  &&  (0.771)  & (0.788)  && (0.468)  & (0.503) \\
[1em]
\hline
Constant       &   4.647$^{***}$ &  4.334$^{***}$   && 4.723$^{***}$ & 4.402$^{***}$  && 4.534$^{***}$ & 4.257$^{***}$ \\
             &   (0.224)   &      (0.276)   &&  (0.250)  & (0.311)  && (0.305)  & (0.349) \\
\hline
Without \textsc{Stable} & & \checkmark &&  & \checkmark &&  & \checkmark \\
Period dummies & \checkmark & \checkmark && \checkmark & \checkmark && \checkmark & \checkmark \\
High Earners dummy & \checkmark & \checkmark && & && & \\

\hline
\(N\)     &     455    &   364  &&   240     & 192    &&   215   & 172  \\
\hline\hline
\end{tabular}

\begin{tablenotes}
\item \footnotesize Note: Giving is indicated in Euro. Any decrease denotes \textsc{IntraDecrease} and \textsc{CatchingUp} for High Earners, and \textsc{IntraDecrease} and \textsc{IntraInterDecrease} for Low Earners. All treatments are included in the regression, except when we indicate that we remove the \textsc{Stable} treatment. Standard errors are in parentheses. $^{*}$ \textit{p}$<$0.10, $^{**}$ \textit{p}$<$0.05, $^{***}$ \textit{p}$<$0.01.
\end{tablenotes}
\end{threeparttable}
}	
\end{table}

\subsection*{Evidence that Misunderstandings do not Drive our Results}
\label{subsec:results_misunderstandings}

Our results are robust to dropping the rare individuals who give almost everything during a period (i.e., $\geq$ 9 Euro from the 11-Euro joint account)---resulting in  a greater income for the participant they are matched with than for themselves---because this might indicate that they misunderstood the instructions (alternatively, those participants could simply be very generous). Appendix Table~\ref{tab:results_withoutfirstperiod} reproduces some of the regressions from Table~\ref{tab:results} for participants with social preferences, but excluding the 2 socially-inclined participants giving 9 Euro or more in at least one period. The results are very similar in terms of significance levels. The exception is that, for all Earners, Wald tests indicate that the difference between any inter decrease and any strict intra decrease is no longer marginally significant (columns 3-4).

\section{Conclusion}
\label{sec:conclusion}

In this study, we conducted an experiment designed to identify how income decreases causally affect redistribution behavior. Based on the literature on reference dependence, loss aversion, and inequality aversion, we hypothesized that individuals become more selfish when they experience a negative absolute or relative income trend. In the full sample, we found evidence that individuals indeed share less with others after experiencing decreasing incomes, but the effect is only significant at marginal levels.

However, since we study the effect of aversion to decreasing income trends on generosity, we should only expect to observe effects for those individuals who are also inequity averse in outcomes. That is, those who are already fully selfish cannot be made more selfish by manipulating income trends. We indeed found that, for those socially-inclined individuals, a negative income trend reduces generosity, this time at highly significant levels, and independently of the specific cutoff that we employ to define those participants. When we conduct separate analyses by role in the experiment, the effect is significant for those participants who earn relatively more, but, in our robust specification, it is not for those who earn relatively less. Furthermore, inter-personal decreases significantly reduce the generosity of participants, and strictly intra-personal decreases generally do not lower generosity significantly. On the whole, inter-personal decreases appear to have a stronger effect than intra-personal decreases, but the difference is only marginally significant, so that we cannot clearly distinguish the two effects.

Our model can be seen as a first step towards a multi-period extension of \citet{bolton2000erc}, which, in the words of its authors “is a theory of ``local behavior" in
the sense that it explains stationary patterns [...] over a short
time span in a constant frame." Specifically, we incorporate inequality aversion in income as well as inequality aversion in income trends. In the fashion of \citet{kHoszegi2006model}, we also assume that individuals are averse to their own trend being negative. Our model reduces to a \citet{bolton2000erc} inequality-aversion model when individuals experience the same positive trends and to a \citet{kHoszegi2006model} reference-dependence model when individuals experience the same negative trends and income levels. When income trends diverge, our theory differs from \citet{bolton2000erc} in that it prescribes a behavior that we observe in the data: a negative relative trend induces more selfishness.\footnote{Conversely, since we model aversion to relative trend inequality similarly to aversion to relative income inequality, our theory also suggests that a positive relative trend induces generosity. This simplifies our theory, but is not supported by the data in our experiment.} For the purpose of this paper, we apply our theory to a dictator game, but natural extensions are to study how  income trends may affect behavior in other settings such as bargaining problems, risky decisions, and consumer decisions.

The effect of decreasing  incomes on generosity that we report also poses interesting possibilities for future research in domains where social preferences are at play. Redistributive policies aiming to close earnings gaps between groups---such as between ethnic minorities vs. whites, and men vs. women---can be met with more resistance by members of the traditionally better off group who are averse to relative income trends inequality and compare their trends to that of the other groups. Aversion to a negative absolute trend can increase the rich's opposition to and the poor's support for fiscal stimuli with redistributive aspects during economic downturns. In the workplace, trend-inequality averse workers might react badly---becoming less productive, less cooperative, engaging in sabotage---when their co-workers face relatively better income profiles over the years.

\newpage
\bibliographystyle{apalike}
\bibliography{income}

\newpage
\appendix
\renewcommand\thefigure{\thesection.\arabic{figure}}    
\renewcommand\thetable{\thesection.\arabic{table}}
\setcounter{table}{0}  \setcounter{figure}{0}



# A   Instructions

<div align="center">

## Instructions

</div>

Welcome to this economic experiment. Please read these instructions carefully. In the experiment, your decisions and the decisions of other participants will determine how much money you earn. You will be paid in cash at the end of the session, provided that you follow the rules. If you do not follow the rules, you will not be paid. You are forbidden from using your phone and from communicating with other participants at any time during the session. If you have any questions, please do not hesitate to raise your hand to ask the experimenter for help.

The **session lasts for up to 90 minutes**. You need to stay until the end to be paid. To complete the task in this experiment, you are also **required to be able to click many times with a computer mouse** (e.g., you need to have no injuries to your arms and fingers).

<div align="center">

### General Instructions

</div>

There are **5 periods, each lasting 8 minutes. Each period, you are matched with 1 anonymous other person**. This can be any participant in the session. **All participants work on the exact same task**. Each **period consists of 2 sub-periods of 4 minutes**. In each sub-period, you are asked to complete a **task to earn a wage**. You need to complete the task in each of the 2 sub-periods in a period to earn the 2 wages for the period. Over the 2 sub-periods, you observe your wage and the wage of the other participant. Your wage and the wage of the other participant may vary over the 2 sub-periods, but the task stays the same.

In each of the 2 sub-periods, a **tax is deducted from your wage and the wage of the other participant** for this period. The tax is 1/3 of the wage. You are left with the after-tax wage (wage minus tax). At the end of the period, the taxes that are taken from you and the other participant over the 2 sub-periods are joined together into the **Total Tax Collected**.

At the end of each of the 5 periods, you **decide how to distribute the Total Tax Collected between you and the other participant**. The **other participant also decides** how to distribute the Total Tax Collected. The decisions are anonymous. Either your distribution or the distribution of the other participant is randomly chosen to count for payment. If you do not complete the task for 1 of the 2 sub-periods for a period, then you do not earn anything in this period and you cannot choose a distribution in this period. The tax taken from the other participant is returned to him or her and there are no distribution decisions.

Until the end of the experiment, you do not receive feedback regarding what distribution the other participant chooses, and the other participant is not told what distribution you choose. Only at the end of the experiment, **1 of the 5 periods is randomly chosen** to count for payment by the computer.

You are paid the following two parts for the chosen period. First, you are **paid the 2 after-tax wages you earned** in that period. Second, either **your distribution of the Total Tax Collected or the distribution of the other participant** you were matched with for this period is **randomly chosen** to be paid.

You need to answer a few **comprehension questions** and go through a **Practice period** before you start the experiment. In the next two sections, you are given more details regarding specific components of

<div align="center">

1

</div>

the experiment, and examples of how your wage and the wage of the other participant are presented during the experiment. Finally, a **timeline** summarizing the different steps is presented at the end of these instructions.

<div align="center">**Details**</div>

**Other Participant:** The other participant **is real and always works on the exact same task as you**. He or she is someone else in the session. At the start of each period, you are matched with a participant for this period. That is, you are matched with the same participant for the 2 sub-periods of this period.

**Periods:** There are 5 periods, each lasting 8 minutes. Between every period, you have a 2-minute break to relax.

**Sub-periods:** Within every period, there are 2 sub-periods of 4 minutes each.

**Task:** The **task is always exactly the same for every participant** in every sub-period. It is a simple task that can be completed by exerting a reasonable effort. The task is to repeatedly click on a ball that appears on your computer screen. Every time you click on the ball, it decreases in size until it disappears. Click on the ball, and wait for it to move until you click again. If you click multiple times before it moves, the ball does not disappear faster. The minimum number of balls that need to disappear (**Balls Threshold**) in each sub-period to earn your wage is **4 balls**. You need to reach this Ball Threshold in each of the 2 sub-periods of a period to earn the 2 wages for the period. This means that if you complete the task in only one of the 2 sub-periods of a period, you do not earn any of the 2 wages for the period. If you want, **you can make more than 4 balls disappear, but this will not change your wage**. That is, you cannot increase your wage by working more than meeting the Balls Threshold.

**Wage and Tax**: In a period, you earn two wages for completing the task: one wage in each of the two sub-periods. The wage may not be the same in the first sub-period and the second sub-period. **If you do not complete the task in each of the 2 sub-periods, you are not paid at all for this period.** That is, you neither earn the wage of the first sub-period nor the wage of the second sub-period. In each sub-period, a tax of 1/3 is removed from your wage and from the wage of the other participant. The tax always amounts to 1/3 of the wage, both for you and the other participant. You are left with the **after-tax wage** (wage minus tax).

**Total Tax Collected:** Taxes collected from you and from the other participant in the 2 sub-periods are added at the end of the period into the **Total Tax Collected**.

**Distribution:** At the end of each period, you choose how to distribute the money in the Total Tax Collected between yourself and the other participant. The other participant also decides how to distribute the Total Tax Collected. Either your distribution or the distribution chosen by the other participant is randomly chosen for payment, for the period randomly selected to be paid out. If you do not complete the task in sub-period 1 or 2 of a period, you cannot choose a distribution. In this case, the tax taken from the other participant is returned to him or her, and there are no distribution decisions. Similarly, if the other participant does not complete the task in one or both of the 2 sub-periods, you and the other

<div align="center">2</div>

participant do not make a distribution choice, and the tax collected on your wages is returned to you. The distribution decisions are anonymous and you are not informed about any of them before the end of the session.

**Payment:** At the end of the session, the computer randomly selects 1 period for payment. There are 2 parts to your payment, provided that you have completed the task in the 2 sub-periods of the chosen period. First, you receive your after-tax wage for each of the 2 sub-periods. Second, your distribution or the distribution of the other participant you were matched with in this period is randomly chosen and paid out. The randomly selected distribution is the only one you are informed about during the experiment. **If you do not complete the task in each the 2 sub-periods of this period, you are paid nothing.**

**Comprehension Questions:** You are asked a few comprehension questions before the experiment starts to make sure that you understand the instructions. If you do not understand a comprehension question, please ask the experimenter for help by raising your hand.

**Practice Period:** During the Practice period, you try the task and become familiar with the experiment, including the distribution decision. The **practice period is shorter than the regular periods, and does not count for payment**. The experiment starts after the practice period.

**Questions:** If anything is unclear, please do not hesitate to raise your hand to ask the experimenter.

3

**Examples**

The following are three examples of how your wage and the wage of the other participant you are matched with for a period are indicated in each of the 2 sub-periods of a period.

**Example 1:** You are matched with a participant for the period. In sub-period 1, your wage is 3 EUR and the wage of the other participant is 4.5 EUR:

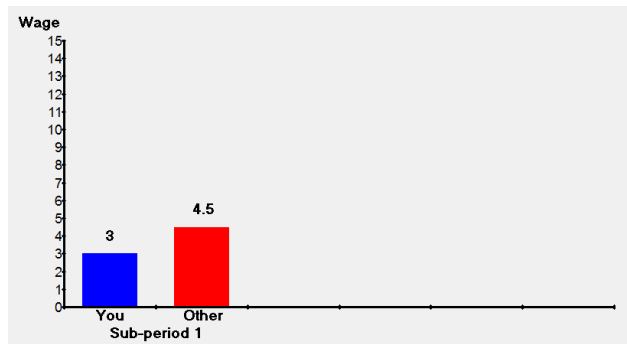

Then, in sub-period 2, your wage increases to 12 EUR and the other participant's wage increases to 13.5 EUR:

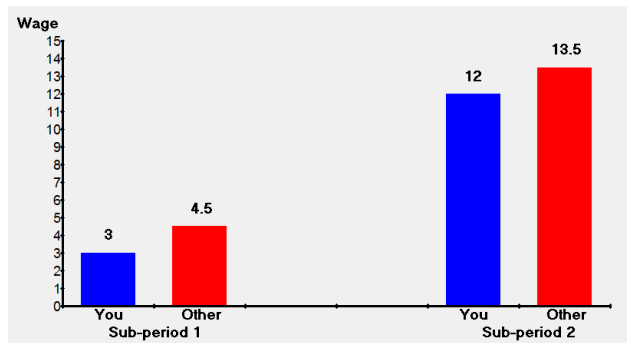

4

**Example 2:** You are matched with a participant for the period. In sub-period 1, your wage is 12 EUR and the wage of the other participant is 13.5 EUR:

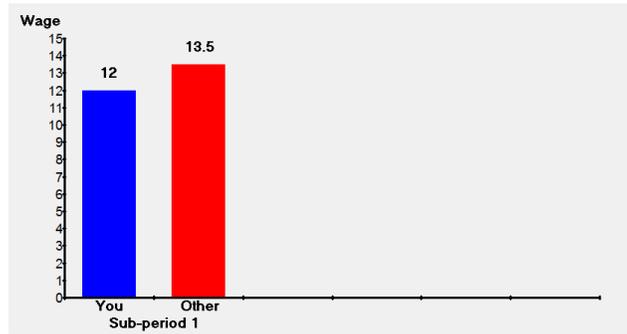

Then, in sub-period 2, your wage decreases to 3 EUR and the other participant's wage decreases to 4.5 EUR:

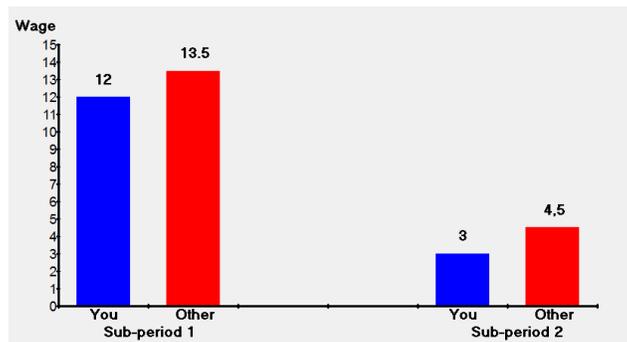

**Example 3:** You are matched with a participant for the period. In sub-period 1, your wage is 6 EUR and the wage of the other participant is also 6 EUR:

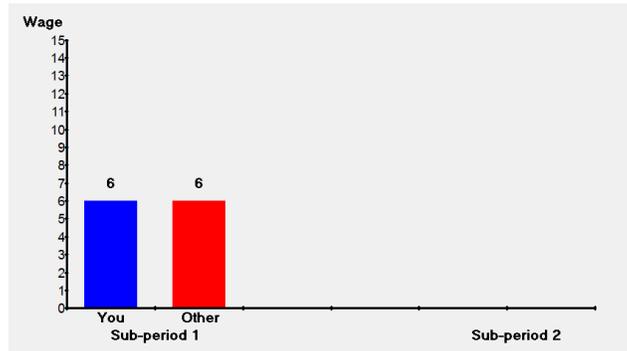

Then, in sub-period 2, your wage stays 6 EUR and the other participant's wage also stays 6 EUR:

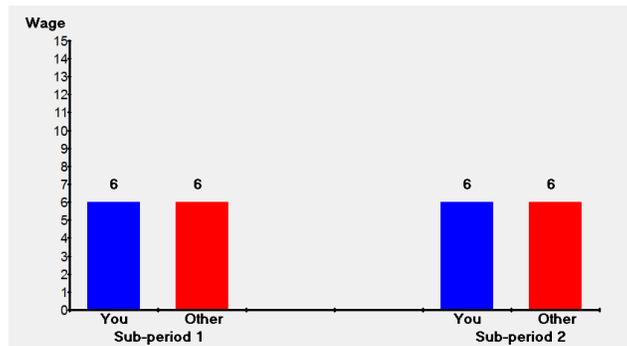

**Timeline**

The following timeline summarizes the different steps in this experiment.

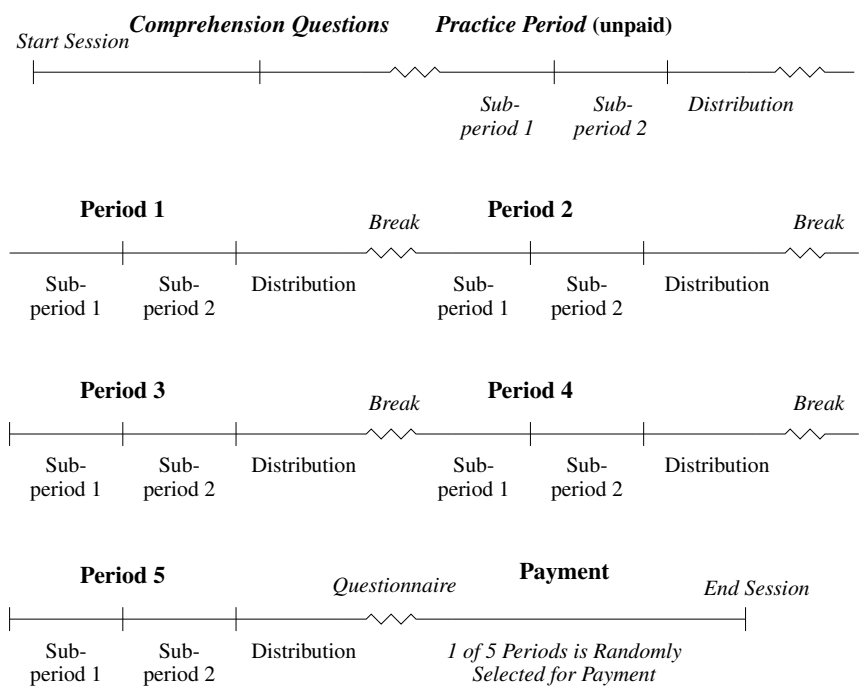

We call participants one by one at the end of the session to pay them. Please stay seated in your cubicle while you wait.



\section{Screenshots}
\label{APP:B}
\begin{figure}[h!]
\centering 
\includegraphics[width=\textwidth]{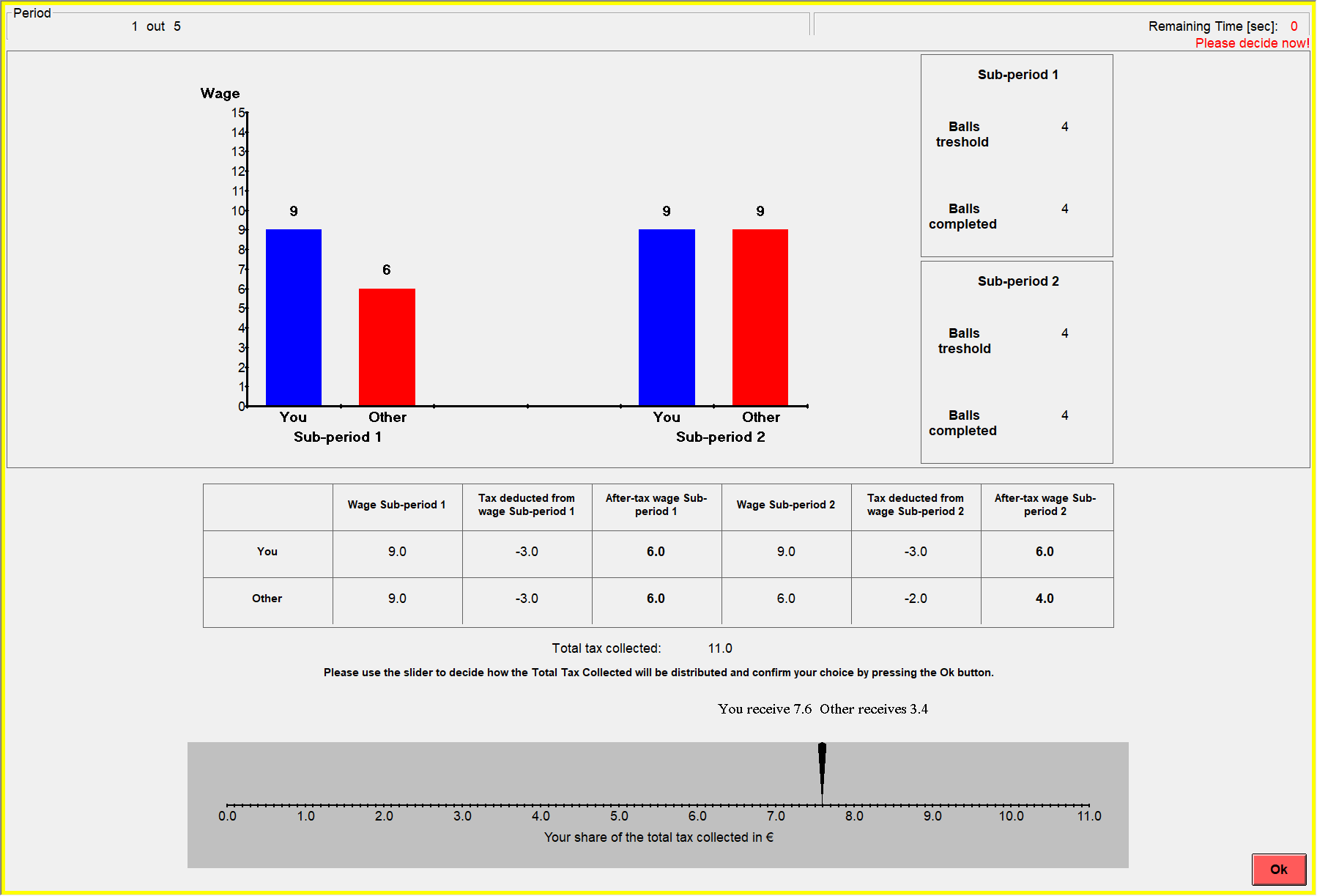}
 \caption{Screenshot of redistribution stage (\textsc{CatchingUp}} treatment)\label{fig:example_redistribution}
\end{figure}

\newpage

\section{Additional Information about the Model}\label{APP:C}
\setcounter{table}{0}  \setcounter{figure}{0}

\noindent{\bf Graphical examples of the dictator' utility function}

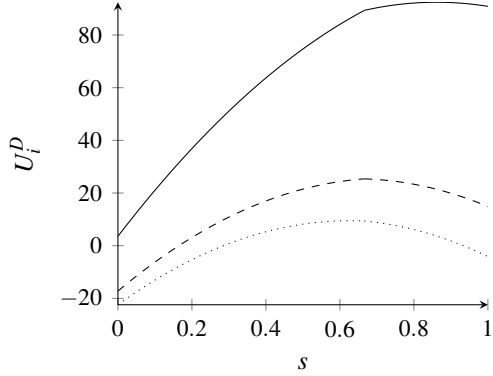
\begin{figure}[ht]
  \centering

\begin{tikzpicture}
\begin{axis}[
    axis lines = left,
    xlabel = $s$,
    ylabel = {$U^D_i$},
]
\addplot [
    domain=0.666:1, 
    samples=100, 
    color=black,
]
{6*(10+9*x)-(0.5)*(((0.5)*(18*x-14))^2+((0.5)*(18*x-5))^2)};

\addplot [
    domain=0:0.666, 
    samples=100, 
    color=black,
    ]
    {6*(10+9*x+(0.8)*(-6+9*x))-(0.5)*(((0.5)*(18*x-14))^2+((0.5)*(18*x-5))^2)};

\addplot [dashed,
    domain=0.666:1, 
    samples=100, 
    color=black,
]
{2*(10+9*x)-(0.5)*(((0.5)*(18*x-14))^2+((0.5)*(18*x-5))^2)};

\addplot [dashed,
    domain=0:0.666, 
    samples=100, 
    color=black,
    ]
    {2*(10+9*x+(0.8)*(-6+9*x))-(0.5)*(((0.5)*(18*x-14))^2+((0.5)*(18*x-5))^2)};

\addplot [dotted,
    domain=0.666:1, 
    samples=100, 
    color=black,
]
{1*(10+9*x)-(0.5)*(((0.5)*(18*x-14))^2+((0.5)*(18*x-5))^2)};

\addplot [dotted,
    domain=0:0.666, 
    samples=100, 
    color=black,
    ]
    {1*(10+9*x+(0.8)*(-6+9*x))-(0.5)*(((0.5)*(18*x-14))^2+((0.5)*(18*x-5))^2)};

\end{axis}
\end{tikzpicture}
    \caption{Examples of $U_i^D$  when $X_1=10, X_2=15, T=9, \Delta_i=-6, \Delta_j=-10$, and 
$\eta=0.8$. Thick, dashed, and dotted curves represent agent types ($a_{i}/b_{i}$) $6$, $2$ and $1$.}
\label{fig:example.add.utility}

\end{figure}

Figure~\ref{fig:example.add.utility} presents examples of a dictator's utility, $U^D_i$, for different agent types ($a_{i}/b_{i}$). 

\medskip

\noindent {\bf Proof of Proposition 1:} Let $\overline{U}_i(y_i,y_j,t_i,t_j)$ and $\hat{U}_i(y_i,y_j,t_i,t_j)$ denote the two piece-wise components of the dictator's utility function $U_i^D(y_i,y_j,t_i,t_j)$. That is:

\begin{align*}
      \hat{U}_i(.)=   a_i(W_i+sT+\eta( \Delta_i+sT)) -b_i\left[\left(\frac{1}{2}(W_i-W_j+2sT-T)\right)^2 + \left(\frac{1}{2}(\Delta_i-\Delta_j+2sT-T)\right)^2
    \right]
        \end{align*}
    \begin{align*}
   \overline{U}_i(.)=     
a_i(W_i+sT)
  -b_i\left[\left(\frac{1}{2}(W_i-W_j+2sT-T)\right)^2 + \left(\frac{1}{2}(\Delta_i-\Delta_j+2sT-T)\right)^2
    \right] 
\end{align*}

Note that $\overline{U}_i(.)$ and $\hat{U}_i(.)$ are both continuous and concave  with unique {\it maxima} at 
\begin{align*}
    \argmax\overline{U}_i(y_i,y_j,t_i,t_j)=\frac{a_i}{b_i}\frac{1}{4T}+\frac{1}{2}+\frac{(X_2-X_1)}{4T}+\frac{(\Delta_j-\Delta_i)}{4T}
\end{align*}
and
\begin{align*}
     \argmax\hat{U}_i(y_i,y_j,t_i,t_j)=\frac{a_i}{b_i}\frac{(1+\eta)}{4T}+\frac{1}{2}+\frac{(X_2-X_1)}{4T}+\frac{(\Delta_j-\Delta_i)}{4T}
\end{align*}
Also note that the piecewise function $U_i^D(y_i,y_j,t_i,t_j)$ has a non differentiable point at $\overline{s}\equiv -\frac{\Delta_i}{T}$.
Because $\hat{U}_i'(\overline{s})>\overline{U}_i'(\overline{s})$, it holds that $U_i^D(y_i,y_j,t_i,t_j)$ is a concave  function as well and thus has a unique maximum.
We consider the following three cases:
\begin{description}

\item[{\bf Case (i)}]  
$ \argmax\overline{U}_i\in [\overline{s}, \infty)$. 
Note that, by definition of $U^D_i$,  for all $s<\overline{s}$ it holds that $\overline{U}_i(s)>\hat{U}_i(s)=U^D_i(s)$. 
It follows that $ \argmax\overline{U}_i=\argmax U^D_i$ whenever $s\geq \overline{s}$.
Recall that, by assumption, $ \argmax\overline{U}_i\geq  \overline{s}$, that is,

\begin{align*}
   \frac{a_i}{b_i}\frac{1}{4T}+\frac{1}{2}+\frac{(X_2-X_1)}{4T}+\frac{(\Delta_j-\Delta_i)}{4T}\geq  -\frac{\Delta_i}{T} \Longleftrightarrow \\
  \frac{a_i}{b_i}\geq  W_i-W_j-2T-3\Delta_i-\Delta_j
\end{align*}

\item[{\bf Case (ii)}]
$ \argmax\hat{U}_i\in (-\infty, \overline{s}]$. 
Note that, by definition of $U^D_i$,  for all $s> \overline{s}$, it holds that $\hat{U}_i(s)>\overline{U}_i(s)=U^D_i(s)$. 
It immediately follows  that $ \argmax\hat{U}_i=\argmax U^D_i$ whenever $s\leq \overline{s}$.
Recall that, by assumption, $ \argmax\overline{U}_i\leq  \overline{s}$, that is,

\begin{align*}
 \frac{a_i}{b_i}\frac{(1+\eta)}{4T}+\frac{1}{2}+\frac{(X_2-X_1)}{4T}+\frac{(\Delta_j-\Delta_i)}{4T}\leq-\frac{\Delta_i}{T} \Longleftrightarrow \\
  \frac{a_i}{b_i}\leq \frac{W_i-W_j-2T-3\Delta_i-\Delta_j}{1+\eta}
\end{align*}

\item[{\bf Case (iii)}]
 $\argmax\hat{U}_i\notin (-\infty, \overline{s})$ and  $\argmax\overline{U}_i\notin [\overline{s}, \infty)$. Since  $\hat{U}_i$ and $\overline{U}_i$ are defined in $U^D_i$ respectively for $s<\overline{s}$ and $s\geq \overline{s}$, it follows that  $\argmax U^D_i=\overline{s}$. Finally, note that, we are in the case such that $\argmax\hat{U}_i>\overline{s}$ and $\argmax\overline{U}_i<\overline{s}$. Thus it follows that  
 \begin{align*}
 \frac{a_i}{b_i}\frac{(1+\eta)}{4T}+\frac{1}{2}+\frac{(X_2-X_1)}{4T}+\frac{(\Delta_j-\Delta_i)}{4T}>-\frac{\Delta_i}{T} \Longleftrightarrow \\
  \frac{a_i}{b_i}> \frac{W_i-W_j-2T-3\Delta_i-\Delta_j}{1+\eta}
\end{align*}
and
\begin{align*}
   \frac{a_i}{b_i}\frac{1}{4T}+\frac{1}{2}+\frac{(X_2-X_1)}{4T}+\frac{(\Delta_j-\Delta_i)}{4T}<  -\frac{\Delta_i}{T} \Longleftrightarrow \\
  \frac{a_i}{b_i}<  W_i-W_j-2T-3\Delta_i-\Delta_j
\end{align*}

\end{description}
Cases (i), (ii), and (iii) together prove the proposition.
$\hfill \blacksquare$

\medskip
\noindent {\bf Proof of Proposition 2:}
Assume that $\frac{a_i}{b_i}\geq \mathbb{U}$. Taking $\mathbb{U}>\mathbb{H}$ together with Proposition 1, it holds that 
\begin{align*}
    \argmax U_i^D=s^*=\frac{a_i}{b_i}\frac{1}{4T}+\frac{1}{2}+\frac{(X_2-X_1)}{4T}+\frac{(\Delta_j-\Delta_i)}{4T}
\end{align*}
To obtain a contradiction, suppose that $s^*<1$. This implies that 
\begin{align*}
    \frac{a_i}{b_i}\frac{1}{4T}+\frac{1}{2}+\frac{(X_2-X_1)}{4T}+\frac{(\Delta_j-\Delta_i)}{4T}<1\Longleftrightarrow\\
    \frac{a_i}{b_i}>W_i-W_j+\Delta_i-\Delta_j+2T=\mathbb{U}
\end{align*}
which is not possible because of the assumption. Thus, for $\frac{a_i}{b_i}\geq \mathbb{U}$, we know that $s^*=1$. 

Next, let's assume that $\frac{a_i}{b_i}\leq \mathbb{L}$. We consider two cases: $
\Delta_1\leq0$ and $\Delta_i>0$. In the former case, it holds that $\mathbb{L}\leq \mathbb{H}(1+\eta)^{-1}$. Therefore, by Proposition 1, it also holds that 
\begin{align*}
      \argmax U_i^D= s^*=\frac{a_i}{b_i}\frac{1+\eta}{4T}+\frac{1}{2}+\frac{(X_2-X_1)}{4T}+\frac{(\Delta_j-\Delta_i)}{4T} 
\end{align*}
To obtain a contradiction, let's assume that $s^*>0$. It follows that
\begin{align*}
    \frac{a_i}{b_i}\frac{1+\eta}{4T}+\frac{1}{2}+\frac{(X_2-X_1)}{4T}+\frac{(\Delta_j-\Delta_i)}{4T}>0\Longleftrightarrow\\
    \frac{a_i}{b_i}>(W_i-W_j+\Delta_i-\Delta_j+2T)(1+\eta)^{-1}=\mathbb{U}>\mathbb{L}
\end{align*}
which is not possible because of the assumption. 
In the latter case, it holds that $\mathbb{L}> \mathbb{H}(1+\eta)^{-1}$, which corresponds to the remaining case 
\begin{align*}
    \argmax U_i^D= s^*=\frac{-\Delta_i}{T}
\end{align*}
Since in this case $\Delta_i>0$, $s^*$ can never be strictly positive.
Finally, given the above, it is straightforward to show that the intermediate case yields $s^*\in (0,1)$.
$\hfill \blacksquare$
\\

\noindent {\bf Proof of Proposition 3:}
Recall that by Proposition 1, $s^*$ depends on the agent type $a_i/b_i$.
For the high and low type ($a_i/b_i>\mathbb{H}$,$a_i/b_i<\mathbb{H}/(1+\eta)$), the value of $s^*$ doe not depend  on $\Delta_i$ whenever $\Delta_i=\Delta_j$. For the intermediate case, it holds that $    \frac{\partial s^*}{\partial \Delta_i}<0$.
$\hfill \blacksquare$
\\

\noindent {\bf Proof of Proposition 4:}
For the high and low type ($a_i/b_i>\mathbb{H}$,$a_i/b_i<\mathbb{H}/(1+\eta)$), it holds that 
$    \frac{\partial s^*}{\partial \Delta_i}>0$. For the intermediate case, the value of $s^*$ does not depends on $\Delta_j$.
$\hfill \blacksquare$

\newpage

\section{Additional Tables}\label{APP:D}
\setcounter{table}{0}  \setcounter{figure}{0}

\begin{table}[h!]	
\centering		
\adjustbox{max width=\textwidth}{

    \begin{threeparttable}
\caption{Summary statistics with different cutoffs to determine participants with sufficiently-strong social preferences}\label{tab:sum_othercutoffs}
\begin{tabular}{rcccccccc}								
\toprule								
\toprule	
 &\multicolumn{8}{c}{Amount given, according to social preferences cutoff (minimum giving in \textsc{Stable})}\\ \cline{2-9}
  &\multicolumn{2}{c}{$\geq$3 Euro} & &\multicolumn{2}{c}{$\geq$1 Euro} & &\multicolumn{2}{c}{$\geq$0.25 Euro}\\ \cline{2-3} \cline{5-6} \cline{8-9}

 Treatment   &High Earners  & Low Earners & &High Earners  &Low Earners & &High Earners  &Low Earners \\
 &	Mean (SD)	&	Mean 	(SD)  & &	Mean (SD)	&	Mean 	(SD)  & &	Mean 	(SD) \\
\midrule							
\textsc{Stable}	& 4.98	& 4.91	& & 4.16 &	3.48 & & 4.09 & 3.21	\\
	& (1.29) &	(1.40) & & (1.91) & (2.09) & & (1.96) & (2.17)	\\
[.2em]
\textsc{IntraDecrease}	& 4.10	& 4.65	& & 3.43 &	3.21 & & 3.38 & 2.97	\\
	& (2.08) &	(1.62) & & (2.33) & (2.25) & & (2.34) & (2.28)	\\
[.2em]
\textsc{IntraInterChange}	& 4.36	&  4.10	& & 3.65 &	2.97 & & 3.60 &  2.75	\\
	& (1.94) &	(1.85) & & (2.22) & (2.09) & & (2.23) & (2.12)	\\
[.2em]
\textsc{CatchingUp}	& 3.78	& 4.55	& & 3.19 &	3.38 & & 3.15 & 3.11	\\
	& (1.94) &	(1.75) & & (2.10) & (2.10) & & (2.10) & (2.18)	\\
[.2em]
\textsc{IntraIncrease}	& 4.59	& 4.38	& & 3.77 &	3.16 & & 3.72 & 2.93	\\
	& (1.66) &	(1.92) & & (2.17) & (2.23) & & (2.18) & (2.26)	\\
	[.2em]
\bottomrule								
\(N\)       & 44 & 37 & & 57 & 62 & & 58 & 68   \\

\bottomrule								
\bottomrule								
\end{tabular}	

\begin{tablenotes}
\item \footnotesize Note: Participants could give any amount between 0 and 11 Euro from the 11-Euro joint account. The remaining amount was credited to their own account. Participants with social preferences are those exhibiting sufficiently-strong advantageous inequality aversion, i.e., they give at least the amount of the cutoff in the baseline treatment \textsc{Stable}. The main text uses 2 Euro as the cutoff and we present here descriptive statistics for the alternative cutoffs 3 Euro, 1 Euro, and 0.25 Euro.
\end{tablenotes}
\end{threeparttable}
}
\end{table}

\begin{table}[h!]							
\centering	
\adjustbox{max width=\textwidth}{
\begin{threeparttable}
\caption{Tests of the effect of decreasing wages on giving, using Fixed Effects instead of Random Effects for Participants with Social Preferences}\label{tab:results_fe}

\begin{tabular}{{r}c*{17}{c}}
\hline\hline
            &\multicolumn{1}{c}{(1)}&\multicolumn{1}{c}{(2)}&  \multicolumn{1}{c}{(3)}&  \multicolumn{1}{c}{(4)}&&\multicolumn{1}{c}{(5)}&\multicolumn{1}{c}{(6)}&\multicolumn{1}{c}{(7)}&\multicolumn{1}{c}{(8)}&&\multicolumn{1}{c}{(9)}&\multicolumn{1}{c}{(10)}&\multicolumn{1}{c}{(11)}&\multicolumn{1}{c}{(12)}&\multicolumn{1}{c}{(13)}&\multicolumn{1}{c}{(14)}\\
            & FE & FE & FE & FE && FE & FE & FE & FE && FE & FE & FE & FE & FE & FE \\
[.5em]
            & \multicolumn{15}{c}{Participants with Social Preferences} \\ \cline{2-17}
 & \multicolumn{4}{c}{All Earners} & & \multicolumn{4}{c}{High Earners} & & \multicolumn{6}{c}{Low Earners} \\ \cline{2-5} \cline{7-10} \cline{12-17} 

&  &  &  &  &&  &  &  &  &&  &  &  &  &  &  \\
Any decrease    & $-$0.483$^{***}$ &      $-$0.332$^{***}$ &   &   &&     $-$0.636$^{***}$&       $-$0.509$^{**}$ &               &               &&      $-$0.322$^{***}$&      $-$0.153   &               &               &               &                \\
            &  (0.117)   &     (0.122)   &   &   &&     (0.200)   &      (0.211)   &               &               &&    (0.115)   &     (0.120)   &               &               &               &                \\
[1em]

Any strictly intra decrease &  &  & $-$0.301$^{**}$ & $-$0.160 &&  &  &  &  &&  &  &  &  &  &  \\
&  &  & (0.118) & (0.134) &&  &  &  &  &&  &  &  &  &  &  \\
[1em]
Any inter decrease &  &  & $-$0.660$^{***}$ & $-$0.502$^{***}$ &&  &  &  &  &&  &  &  &  &  &  \\
&  &  & (0.172) & (0.166) &&  &  &  &  &&  &  &  &  &  &  \\
[1em]
\hline
Wald test \textit{p}-value & & & 0.053 & 0.058 && & & & && & & & & & \\
\hline
&  &  &  &  &&  &  &  &  &&  &  &  &  &  &  \\
\textsc{IntraDecrease}      &                &               &    &    &&              &                &      $-$0.492$^{***}$&      $-$0.367$^{*}$  &&              &               &     $-$0.094   &      0.064   &               &                \\
            &              &               &    &    &&              &                &     (0.182)   &     (0.203)   &&              &               &     (0.135)   &     (0.167)   &               &                \\
[1em]
\textsc{CatchingUp}      &            &              &    &    &&               &                &      $-$0.781$^{**}$ &      $-$0.657$^{**}$ &&              &               &               &               &               &                \\
            &                  &               &    &    &&              &                &     (0.308)   &     (0.310)   &&              &               &               &               &               &                \\
[1em]
\textsc{IntraInterDecrease}      &          &               &    &    &&              &                &               &               &&              &               &      $-$0.551$^{***}$&      $-$0.377$^{**}$ &      $-$0.527$^{***}$&      $-$0.399$^{**}$  \\
            &               &               &    &    &&              &                &               &               &&              &               &     (0.181)   &     (0.164)   &     (0.182)   &     (0.169)    \\
[1em]
\hline
Wald test \textit{p}-value & & & & && & & 0.357 & 0.351 && & & 0.044 & 0.055 & & \\
\hline
Constant      &     4.475$^{***}$ &       4.205$^{***}$ &  4.482$^{***}$  &  4.215$^{***}$  &&      4.556$^{***}$ &        4.228$^{***}$ &       4.547$^{***}$ &       4.216$^{***}$ &&      4.361$^{***}$ &       4.139$^{***}$ &       4.398$^{***}$ &       4.193$^{***}$ &       4.378$^{***}$ &       4.214$^{***}$ \\
            &    (0.144)   &     (0.165)   &  (0.141)  &  (0.160)  &&    (0.227)   &      (0.248)   &     (0.232)   &     (0.254)   &&    (0.163)   &     (0.209)   &     (0.158)   &     (0.203)   &     (0.159)   &     (0.216)    \\
\hline
Without \textsc{Stable} & & \checkmark & & \checkmark && & \checkmark & & \checkmark && & \checkmark & & \checkmark & & \checkmark \\
Period dummies & \checkmark & \checkmark & \checkmark & \checkmark && \checkmark & \checkmark & \checkmark & \checkmark && \checkmark & \checkmark & \checkmark & \checkmark & \checkmark & \\
\hline
\(N\)       &        455   &         364   &  455   &  364   &&        240   &          192   &         240   &         192     &&        215   &         172   &         215   &         172   &         215   &         172    \\
\hline\hline
\end{tabular}

\begin{tablenotes}
\item \footnotesize Note: Giving is indicated in Euro. Participants with social preferences are those exhibiting sufficiently-strong advantageous inequality aversion, i.e., they give at least 2 Euro in the baseline treatment \textsc{Stable}. Any decrease denotes \textsc{IntraDecrease} and \textsc{CatchingUp} for High Earners, and \textsc{IntraDecrease} and \textsc{IntraInterDecrease} for Low Earners. Any strictly intra decrease denotes \textsc{IntraDecrease} for High and Low Earners. Any inter decrease denotes \textsc{CatchingUp} for High Earners and \textsc{IntraInterDecrease} for Low Earners. All treatments are included in the regression, except when we indicate that we remove the \textsc{Stable} treatment. We indicate \textit{p}-values from Wald test on the equality of the coefficients of Any strictly intra decrease and Any inter decrease, and on the equality of \textsc{IntraInterDecrease} and \textsc{IntraDecrease} or \textsc{CatchingUp}. There is no need for a High Earners dummy because FE does not incorporate time-invariant variables. Standard errors are in parentheses. $^{*}$ \textit{p}$<$0.10, $^{**}$ \textit{p}$<$0.05, $^{***}$ \textit{p}$<$0.01.
\end{tablenotes}
\end{threeparttable}
}	
\end{table}

\begin{table}[h!]							
\centering	
\adjustbox{max width=\textwidth}{
\begin{threeparttable}
\caption{Tests of the effect of decreasing wages on giving, using the alternative definition that participants with social preferences are those giving at least 3 Euro in \textsc{Stable}}\label{tab:results_socialpreferences300}

\begin{tabular}{{r}c*{17}{c}}
\hline\hline
            &\multicolumn{1}{c}{(1)}&\multicolumn{1}{c}{(2)}&  \multicolumn{1}{c}{(3)}&  \multicolumn{1}{c}{(4)}&&\multicolumn{1}{c}{(5)}&\multicolumn{1}{c}{(6)}&\multicolumn{1}{c}{(7)}&\multicolumn{1}{c}{(8)}&&\multicolumn{1}{c}{(9)}&\multicolumn{1}{c}{(10)}&\multicolumn{1}{c}{(11)}&\multicolumn{1}{c}{(12)}&\multicolumn{1}{c}{(13)}&\multicolumn{1}{c}{(14)}\\
            & RE & RE & RE & RE && RE & RE & RE & RE && RE & RE & RE & RE & RE & RE \\
[.5em]
            & \multicolumn{15}{c}{Participants with Social Preferences} \\ \cline{2-17}
 & \multicolumn{4}{c}{All Earners} & & \multicolumn{4}{c}{High Earners} & & \multicolumn{6}{c}{Low Earners} \\ \cline{2-5} \cline{7-10} \cline{12-17} 

&  &  &  &  &&  &  &  &  &&  &  &  &  &  &  \\
Any decrease    & $-$0.483$^{***}$ &      $-$0.338$^{**}$ &   &   &&     $-$0.681$^{***}$ &       $-$0.563$^{**}$ &               &               &&      $-$0.263$^{**}$ &      $-$0.069   &               &               &               &                \\
            &  (0.128)   &     (0.134)   &   &   &&     (0.217)   &      (0.228)   &               &               &&    (0.114)   &     (0.105)   &               &               &               &                \\
[1em]

Any strictly intra decrease &  &  & $-$0.250$^{**}$ & $-$0.116 &&  &  &  &  &&  &  &  &  &  &  \\
&  &  & (0.126) & (0.144) &&  &  &  &  &&  &  &  &  &  &  \\
[1em]
Any inter decrease &  &  & $-$0.707$^{***}$ & $-$0.553$^{***}$ &&  &  &  &  &&  &  &  &  &  &  \\
&  &  & (0.189) & (0.180) &&  &  &  &  &&  &  &  &  &  &  \\
[1em]
\hline
Wald test \textit{p}-value & & & 0.019 & 0.021 && & & & && & & & & & \\
\hline
&  &  &  &  &&  &  &  &  &&  &  &  &  &  &  \\
\textsc{IntraDecrease}      &                &               &    &    &&              &                &      $-$0.495$^{**}$&      $-$0.388$^{*}$  &&              &               &     0.028   &      0.202   &               &                \\
            &              &               &    &    &&              &                &     (0.195)   &     (0.223)   &&              &               &     (0.124)   &     (0.155)   &               &                \\
[1em]
\textsc{CatchingUp}      &            &              &    &    &&               &                &      $-$0.868$^{***}$ &      $-$0.747$^{**}$ &&              &               &               &               &               &                \\
            &                  &               &    &    &&              &                &     (0.330)   &     (0.326)   &&              &               &               &               &               &                \\
[1em]
\textsc{IntraInterDecrease}      &          &               &    &    &&              &                &               &               &&              &               &      $-$0.553$^{***}$ &      $-$0.350$^{**}$ &      $-$0.560$^{***}$ &      $-$0.421$^{**}$ \\
            &               &               &    &    &&              &                &               &               &&              &               &     (0.200)   &     (0.171)   &     (0.204)   &     (0.186)    \\
[1em]
\hline
Wald test \textit{p}-value & & & & && & & 0.256 & 0.262 && & & 0.017 & 0.026 & & \\
\hline
Constant      &     4.702$^{***}$ &       4.377$^{***}$ & 4.711$^{***}$   &  4.390$^{***}$  &&  4.791$^{***}$     &  4.468$^{***}$       &   4.778$^{***}$    &   4.451$^{***}$     &&  4.710$^{***}$     &   4.393$^{***}$    &   4.765$^{***}$     &   4.473$^{***}$     &       4.771$^{***}$ &      4.543$^{***}$ \\
            &    (0.174)   &     (0.236)   &  (0.173)  &  (0.233)  &&    (0.181)   &      (0.261)   &   (0.185)   &     (0.266)   &&    (0.226)   &     (0.267)   &     (0.219)   &     (0.256)   &     (0.216)   &     (0.261)    \\
\hline
Without \textsc{Stable} & & \checkmark & & \checkmark && & \checkmark & & \checkmark && & \checkmark & & \checkmark & & \checkmark \\
Period dummies & \checkmark & \checkmark & \checkmark & \checkmark && \checkmark & \checkmark & \checkmark & \checkmark && \checkmark & \checkmark & \checkmark & \checkmark & \checkmark & \\
High Earners dummy & \checkmark & \checkmark & \checkmark & \checkmark && &  & &  && &  & & & & \\

\hline
\(N\)       &        405   &         324   &  405   &  324   &&        220   &          176   &       220    &        176     &&        185   &         148   &        185    &        148   &        185   &         148    \\
\hline\hline
\end{tabular}

\begin{tablenotes}
\item \footnotesize Note: Giving is indicated in Euro. Participants with social preferences are those exhibiting sufficiently-strong advantageous inequality aversion, i.e., they give at least 3 Euro in the baseline treatment \textsc{Stable}. Any decrease denotes \textsc{IntraDecrease} and \textsc{CatchingUp} for High Earners, and \textsc{IntraDecrease} and \textsc{IntraInterDecrease} for Low Earners. Any strictly intra decrease denotes \textsc{IntraDecrease} for High and Low Earners. Any inter decrease denotes \textsc{CatchingUp} for High Earners and \textsc{IntraInterDecrease} for Low Earners. All treatments are included in the regression, except when we indicate that we remove the \textsc{Stable} treatment. We indicate \textit{p}-values from Wald test on the equality of the coefficients of Any strictly intra decrease and Any inter decrease, and on the equality of \textsc{IntraInterDecrease} and \textsc{IntraDecrease} or \textsc{CatchingUp}. The coefficient of the High Earners dummy is always far from the 10\% significance threshold. Standard errors are in parentheses. $^{*}$ \textit{p}$<$0.10, $^{**}$ \textit{p}$<$0.05, $^{***}$ \textit{p}$<$0.01.
\end{tablenotes}
\end{threeparttable}
}	
\end{table}

\begin{table}[h!]							
\centering	
\adjustbox{max width=\textwidth}{
\begin{threeparttable}
\caption{Tests of the effect of decreasing wages on giving, using the alternative definition that participants with social preferences are those giving at least 1 Euro in \textsc{Stable}}\label{tab:results_socialpreferences100}

\begin{tabular}{{r}c*{17}{c}}
\hline\hline
            &\multicolumn{1}{c}{(1)}&\multicolumn{1}{c}{(2)}&  \multicolumn{1}{c}{(3)}&  \multicolumn{1}{c}{(4)}&&\multicolumn{1}{c}{(5)}&\multicolumn{1}{c}{(6)}&\multicolumn{1}{c}{(7)}&\multicolumn{1}{c}{(8)}&&\multicolumn{1}{c}{(9)}&\multicolumn{1}{c}{(10)}&\multicolumn{1}{c}{(11)}&\multicolumn{1}{c}{(12)}&\multicolumn{1}{c}{(13)}&\multicolumn{1}{c}{(14)}\\
            & RE & RE & RE & RE && RE & RE & RE & RE && RE & RE & RE & RE & RE & RE \\
[.5em]
            & \multicolumn{15}{c}{Participants with Social Preferences} \\ \cline{2-17}
 & \multicolumn{4}{c}{All Earners} & & \multicolumn{4}{c}{High Earners} & & \multicolumn{6}{c}{Low Earners} \\ \cline{2-5} \cline{7-10} \cline{12-17} 

&  &  &  &  &&  &  &  &  &&  &  &  &  &  &  \\
Any decrease    & $-$0.376$^{***}$ &      $-$0.280$^{***}$ &   &   &&     $-$0.532$^{***}$ &       $-$0.427$^{**}$ &               &               &&      $-$0.243$^{**}$ &      $-$0.163   &               &               &               &                \\
            &  (0.097)   &     (0.104)   &   &   &&     (0.177)   &      (0.191)   &               &               &&    (0.096)   &     (0.106)   &               &               &               &                \\
[1em]

Any strictly intra decrease &  &  & $-$0.237$^{**}$ & $-$0.146 &&  &  &  &  &&  &  &  &  &  &  \\
&  &  & (0.097) & (0.108) &&  &  &  &  &&  &  &  &  &  &  \\
[1em]
Any inter decrease &  &  & $-$0.513$^{***}$ & $-$0.413$^{***}$ &&  &  &  &  &&  &  &  &  &  &  \\
&  &  & (0.144) & (0.145) &&  &  &  &  &&  &  &  &  &  &  \\
[1em]
\hline
Wald test \textit{p}-value & & & 0.068 & 0.073 && & & & && & & & & & \\
\hline
&  &  &  &  &&  &  &  &  &&  &  &  &  &  &  \\
\textsc{IntraDecrease}      &                &               &    &    &&              &                &      $-$0.387$^{**}$&      $-$0.285  &&              &               &     $-$0.097   &     $-$0.019   &               &                \\
            &              &               &    &    &&              &                &     (0.164)   &     (0.177)   &&              &               &     (0.105)   &     (0.129)   &               &                \\
[1em]
\textsc{CatchingUp}      &            &              &    &    &&               &                &      $-$0.676$^{**}$ &      $-$0.575$^{**}$ &&              &               &               &               &               &                \\
            &                  &               &    &    &&              &                &     (0.264)   &     (0.274)   &&              &               &               &               &               &                \\
[1em]
\textsc{IntraInterDecrease}      &          &               &    &    &&              &                &               &               &&              &               &      $-$0.388$^{**}$ &      $-$0.307$^{**}$ &      $-$0.364$^{**}$ &      $-$0.300$^{**}$  \\
            &               &               &    &    &&              &                &               &               &&              &               &     (0.154)   &     (0.151)   &     (0.155)   &     (0.151)    \\
[1em]
\hline
Wald test \textit{p}-value & & & & && & & 0.268 & 0.259 && & & 0.109 & 0.118 & & \\
\hline
Constant      &     3.974$^{***}$ &       3.745$^{***}$ &  3.979$^{***}$  &  3.751$^{***}$  &&      3.982$^{***}$ &        3.696$^{***}$ &       3.971$^{***}$ &       3.683$^{***}$ &&      3.550$^{***}$ &      3.439$^{***}$ &       3.569$^{***}$ &       3.464$^{***}$ &       3.548$^{***}$ &       3.458$^{***}$ \\
            &    (0.243)   &     (0.267)   &  (0.243)  & (0.266)    &&   (0.258)    &    (0.285)     &    (0.260)    &    (0.288)    &&   (0.244)    &   (0.249)     &   (0.246)     &  (0.250)      &     (0.248)   &     (0.256)    \\
\hline
Without \textsc{Stable} & & \checkmark & & \checkmark && & \checkmark & & \checkmark && & \checkmark & & \checkmark & & \checkmark \\
Period dummies & \checkmark & \checkmark & \checkmark & \checkmark && \checkmark & \checkmark & \checkmark & \checkmark && \checkmark & \checkmark & \checkmark & \checkmark & \checkmark & \\
High Earners dummy & \checkmark & \checkmark & \checkmark & \checkmark && &  & &  && &  & & & & \\

\hline
\(N\)       &        595   &         476   &  595   &  476   &&        285   &          228   &         285   &         228     &&        310   &         248   &         310   &         248   &         310   &         248    \\
\hline\hline
\end{tabular}

\begin{tablenotes}
\item \footnotesize Note: Giving is indicated in Euro. Participants with social preferences are those exhibiting sufficiently-strong advantageous inequality aversion, i.e., they give at least 1 Euro in the baseline treatment \textsc{Stable}. Any decrease denotes \textsc{IntraDecrease} and \textsc{CatchingUp} for High Earners, and \textsc{IntraDecrease} and \textsc{IntraInterDecrease} for Low Earners. Any strictly intra decrease denotes \textsc{IntraDecrease} for High and Low Earners. Any inter decrease denotes \textsc{CatchingUp} for High Earners and \textsc{IntraInterDecrease} for Low Earners. All treatments are included in the regression, except when we indicate that we remove the \textsc{Stable} treatment. We indicate \textit{p}-values from Wald test on the equality of the coefficients of Any strictly intra decrease and Any inter decrease, and on the equality of \textsc{IntraInterDecrease} and \textsc{IntraDecrease} or \textsc{CatchingUp}. The coefficient of the High Earners dummy is always far from the 10\% significance threshold. Standard errors are in parentheses. $^{*}$ \textit{p}$<$0.10, $^{**}$ \textit{p}$<$0.05, $^{***}$ \textit{p}$<$0.01.
\end{tablenotes}
\end{threeparttable}
}	
\end{table}

\begin{table}[h!]							
\centering	
\adjustbox{max width=\textwidth}{
\begin{threeparttable}
\caption{Tests of the effect of decreasing wages on giving, using the alternative definition that participants with social preferences are those giving at least 0.25 Euro in \textsc{Stable}}\label{tab:results_socialpreferences025}

\begin{tabular}{{r}c*{17}{c}}
\hline\hline
            &\multicolumn{1}{c}{(1)}&\multicolumn{1}{c}{(2)}&  \multicolumn{1}{c}{(3)}&  \multicolumn{1}{c}{(4)}&&\multicolumn{1}{c}{(5)}&\multicolumn{1}{c}{(6)}&\multicolumn{1}{c}{(7)}&\multicolumn{1}{c}{(8)}&&\multicolumn{1}{c}{(9)}&\multicolumn{1}{c}{(10)}&\multicolumn{1}{c}{(11)}&\multicolumn{1}{c}{(12)}&\multicolumn{1}{c}{(13)}&\multicolumn{1}{c}{(14)}\\
            & RE & RE & RE & RE && RE & RE & RE & RE && RE & RE & RE & RE & RE & RE \\
[.5em]
            & \multicolumn{15}{c}{Participants with Social Preferences} \\ \cline{2-17}
 & \multicolumn{4}{c}{All Earners} & & \multicolumn{4}{c}{High Earners} & & \multicolumn{6}{c}{Low Earners} \\ \cline{2-5} \cline{7-10} \cline{12-17} 

&  &  &  &  &&  &  &  &  &&  &  &  &  &  &  \\
Any decrease    & $-$0.353$^{***}$ &      $-$0.263$^{***}$ &   &   &&     $-$0.526$^{***}$ &       $-$0.425$^{**}$ &               &               &&      $-$0.214$^{**}$ &      $-$0.143
   &               &               &               &                \\
            &  (0.092)   &     (0.099)   &   &   &&     (0.173)   &      (0.185)   &               &               &&    (0.090)   &     (0.099)   &               &               &               &                \\
[1em]

Any strictly intra decrease &  &  & $-$0.220$^{**}$ & $-$0.135 &&  &  &  &  &&  &  &  &  &  &  \\
&  &  & (0.092) & (0.102) &&  &  &  &  &&  &  &  &  &  &  \\
[1em]
Any inter decrease &  &  & $-$0.484$^{***}$ & $-$0.391$^{***}$ &&  &  &  &  &&  &  &  &  &  &  \\
&  &  & (0.137) & (0.138) &&  &  &  &  &&  &  &  &  &  &  \\
[1em]
\hline
Wald test \textit{p}-value & & & 0.065 & 0.069 && & & & && & & & & & \\
\hline
&  &  &  &  &&  &  &  &  &&  &  &  &  &  &  \\
\textsc{IntraDecrease}      &                &               &    &    &&              &                &      $-$0.389$^{**}$&      $-$0.288$^{*}$  &&              &               &     $-$0.074   &      $-$0.006   &               &                \\
            &              &               &    &    &&              &                &     (0.162)   &     (0.172)   &&              &               &     (0.097)   &     (0.118)   &               &                \\
[1em]
\textsc{CatchingUp}      &            &              &    &    &&               &                &      $-$0.663$^{**}$ &      $-$0.570$^{**}$ &&              &               &               &               &               &                \\
            &                  &               &    &    &&              &                &     (0.259)   &     (0.269)   &&              &               &               &               &               &                \\
[1em]
\textsc{IntraInterDecrease}      &          &               &    &    &&              &                &               &               &&              &               &      $-$0.353$^{**}$ &      $-$0.281$^{**}$ &      $-$0.334$^{**}$ &      $-$0.280$^{**}$  \\
            &               &               &    &    &&              &                &               &               &&              &               &     (0.143)   &     (0.140)   &     (0.142)   &     (0.138)    \\
[1em]
\hline
Wald test \textit{p}-value & & & & && & & 0.291 & 0.271 && & & 0.091 & 0.098 & & \\
\hline
High Earner dummy      &  $-$0.592$^{*}$ &  $-$0.527  &  $-$0.592$^{*}$  &  $-$0.528  &&  &  &  &  &&  &  &  &  &  &  \\
            & (0.354) &  (0.363) &  (0.355)  &  (0.364)  &&  &  &  &  &&  &  &  &  &  &    \\
[1em]
Constant      &     3.916$^{***}$ &       3.697$^{***}$ &  3.919$^{***}$  &  3.702$^{***}$  &&      3.931$^{***}$ &        3.648$^{***}$ &       3.919$^{***}$ &       3.633$^{***}$ &&      3.294$^{***}$ &      3.198$^{***}$ &       3.312$^{***}$ &       3.223$^{***}$ &       3.297$^{***}$ &       3.221$^{***}$ \\
            &    (0.244)   &     (0.265)   &  (0.244)  &  (0.265)  &&    (0.259)   &      (0.284)   &     (0.261)   &     (0.288)   &&    (0.243)   &     (0.246)   &     (0.245)   &     (0.247)   &     (0.246)   &     (0.252)    \\
\hline
Without \textsc{Stable} & & \checkmark & & \checkmark && & \checkmark & & \checkmark && & \checkmark & & \checkmark & & \checkmark \\
Period dummies & \checkmark & \checkmark & \checkmark & \checkmark && \checkmark & \checkmark & \checkmark & \checkmark && \checkmark & \checkmark & \checkmark & \checkmark & \checkmark & \\

\hline
\(N\)       &        630   &         504   &  630   &  504   &&        290   &          232   &         290   &         232     &&        340   &         272   &         340   &         272   &         340   &         272    \\
\hline\hline
\end{tabular}

\begin{tablenotes}
\item \footnotesize Note: Giving is indicated in Euro. Participants with social preferences are those exhibiting sufficiently-strong advantageous inequality aversion, i.e., they give at least 0.25 Euro in the baseline treatment \textsc{Stable}. Any decrease denotes \textsc{IntraDecrease} and \textsc{CatchingUp} for High Earners, and \textsc{IntraDecrease} and \textsc{IntraInterDecrease} for Low Earners. Any strictly intra decrease denotes \textsc{IntraDecrease} for High and Low Earners. Any inter decrease denotes \textsc{CatchingUp} for High Earners and \textsc{IntraInterDecrease} for Low Earners. All treatments are included in the regression, except when we indicate that we remove the \textsc{Stable} treatment. We indicate \textit{p}-values from Wald test on the equality of the coefficients of Any strictly intra decrease and Any inter decrease, and on the equality of \textsc{IntraInterDecrease} and \textsc{IntraDecrease} or \textsc{CatchingUp}. Standard errors are in parentheses. $^{*}$ \textit{p}$<$0.10, $^{**}$ \textit{p}$<$0.05, $^{***}$ \textit{p}$<$0.01.
\end{tablenotes}
\end{threeparttable}
}	
\end{table}

\begin{table}[h!]							
\centering	
\adjustbox{max width=\textwidth}{
\begin{threeparttable}
\caption{Tests of the effect of decreasing wages on giving, excluding participants giving more than 9 Euro}\label{tab:results_withoutfirstperiod}

\begin{tabular}{{r}c*{13}{c}}
\hline\hline
            & \multicolumn{1}{c}{(1)} & \multicolumn{1}{c}{(2)} & \multicolumn{1}{c}{(3)} & \multicolumn{1}{c}{(4)} && \multicolumn{1}{c}{(5)}&\multicolumn{1}{c}{(6)} & \multicolumn{1}{c}{(7)} & \multicolumn{1}{c}{(8)} &&\multicolumn{1}{c}{(9)}&\multicolumn{1}{c}{(10)} & \multicolumn{1}{c}{(11)} & \multicolumn{1}{c}{(12)} \\
            & RE & RE & RE & RE && RE & RE & RE & RE && RE & RE & RE & RE \\
[.5em]
            & \multicolumn{14}{c}{Participants with Social Preferences} \\ \cline{2-15}
 & \multicolumn{4}{c}{All Earners} && \multicolumn{4}{c}{High Earners} && \multicolumn{4}{c}{Low Earners} \\  \cline{2-5} \cline{7-10} \cline{12-15}

&  &  &  &  &&  &  &  &  &&  &  &  & \\
Any decrease    & $-$0.394$^{***}$ & $-$0.245$^{**}$ &  &  && $-$0.504$^{***}$ & $-$0.372$^{**}$ &  &  &&  $-$0.270$^{***}$ & $-$0.108 &  &       \\
             & (0.097) & (0.105) &  &  &&  (0.152)  & (0.166) &  &  && (0.104) & (0.113) &  &                  \\
[1em]
Any strictly intra decrease  &  &  & $-$0.297$^{**}$ & $-$0.152 &&  &  &  &  &&  &  &  &  \\
&  &  & (0.120) & (0.135) &&  &  &  &  &&  &  &  & \\
[1em]
Any inter decrease  &  &  & $-$0.489$^{***}$ & $-$0.337$^{***}$ &&  &  &  &  &&  &  &  &   \\
 &  &  & (0.121) & (0.117) &&  &  &  &  &&  &  &  &   \\
 &  &  &  &  &&  &  &&  & \\
\hline
Wald test \textit{p}-value  &  &  & 0.174 & 0.189 &&  &  &  &  &&  &  &  & \\
\hline
&  &  &  &  &&  &  &  &  &&  &  &  & \\
\textsc{IntraDecrease}  &  &  &  &  &&  &  & $-$0.478$^{***}$ & $-$0.343$^{*}$ &&  &  & $-$0.093 & 0.061 \\
 &  &  &  &  &&  &  & (0.184) & (0.203) &&  &  & (0.137) & (0.167) \\
[1em]
\textsc{CatchingUp}  &  &  &  &  &&  &  & $-$0.530$^{***}$ & $-$0.402$^{**}$ &&  &  &  &  \\
 &  &  &  &  &&  &  & (0.184) & (0.190) &&  &  &  &  \\
[1em]
\textsc{IntraInterDecrease}  &  &  &  &  &&  &  &  &  &&  &  & $-$0.447$^{***}$ & $-$0.282$^{**}$ \\
 &  &  &  &  &&  &  &  &  &&  &  & (0.152) & (0.137) \\
[1em]
\hline
Wald test \textit{p}-value  &  &  &  &  &&  &  & 0.803 & 0.780 &&  & & 0.076 & 0.094 \\
\hline Constant       & 4.503$^{***}$ & 4.260$^{***}$ & 4.505$^{***}$ & 4.262$^{***}$ && 4.605$^{***}$ & 4.333$^{***}$ & 4.603 & 4.330 && 4.302$^{***}$ & 4.109$^{***}$ & 4.327$^{***}$
 & 4.146$^{***}$ \\
             & (0.202) & (0.249) & (0.202) & (0.248) && (0.211) & (0.260) & (0.213) & (0.264) && (0.245) & (0.279) & (0.241) & (0.273) \\
\hline
Exclude giving $>$ 9 Euro & \checkmark & \checkmark & \checkmark & \checkmark && \checkmark & \checkmark & \checkmark & \checkmark && \checkmark & \checkmark & \checkmark & \checkmark \\
Without \textsc{Stable} &  & \checkmark &  & \checkmark &&  & \checkmark &  & \checkmark &&  & \checkmark &  & \checkmark \\
Period dummies & \checkmark & \checkmark & \checkmark & \checkmark && \checkmark & \checkmark & \checkmark & \checkmark && \checkmark & \checkmark & \checkmark & \checkmark \\
High Earners dummy & \checkmark & \checkmark & \checkmark & \checkmark &&  &  &  &  &&  &  &  &  \\

\hline
\(N\)     & 445 & 356 & 445 & 356 && 235 & 188 & 235 & 188 && 210 & 168 & 210 & 168  \\
\hline\hline
\end{tabular}

\begin{tablenotes}
\item \footnotesize Note: Giving is indicated in Euro. Any decrease denotes \textsc{IntraDecrease} and \textsc{CatchingUp} for High Earners, and \textsc{IntraDecrease} and \textsc{IntraInterDecrease} for Low Earners. Any strictly intra decrease denotes \textsc{IntraDecrease} for High and Low Earners. Any inter decrease denotes \textsc{CatchingUp} for High Earners and \textsc{IntraInterDecrease} for Low Earners. All treatments are included in the regression, except when we indicate that we remove the \textsc{Stable} treatment. We indicate \textit{p}-values from Wald test on the equality of the coefficients of Any strictly intra decrease and Any inter decrease, and on the equality of \textsc{IntraInterDecrease} and \textsc{IntraDecrease} or \textsc{CatchingUp}. Standard errors are in parentheses. $^{*}$ \textit{p}$<$0.10, $^{**}$ \textit{p}$<$0.05, $^{***}$ \textit{p}$<$0.01.
\end{tablenotes}
\end{threeparttable}
}	
\end{table}

\end{document}